\documentclass[%
 reprint,
 superscriptaddress,
 amsmath,amssymb,
 aps,
]{revtex4-2}

\usepackage[english]{babel}
\usepackage{graphicx}
\usepackage{dcolumn}
\usepackage{bm}
\usepackage{soul}

\usepackage{xcolor}

\begin{document}

\preprint{APS/123-QED}

\title{Observation of the relativistic reversal of the ponderomotive potential}

\author{Jeremy J. Axelrod}
\affiliation{%
Dept. of Physics, University of California, Berkeley, USA
}%
\affiliation{%
Lawrence Berkeley National Laboratory, Berkeley, USA
}%

\author{Sara L. Campbell}%
\affiliation{%
 Dept. of Physics, University of California, Berkeley, USA
}%
\affiliation{%
Lawrence Berkeley National Laboratory, Berkeley, USA
}%

\author{Osip Schwartz}%
\affiliation{%
 Dept. of Physics, University of California, Berkeley, USA
}%
\affiliation{%
Lawrence Berkeley National Laboratory, Berkeley, USA
}%

\author{Carter Turnbaugh}%
\affiliation{%
 Dept. of Physics, University of California, Berkeley, USA
}%
\affiliation{%
Lawrence Berkeley National Laboratory, Berkeley, USA
}%

\author{Robert M. Glaeser}%

\affiliation{%
Lawrence Berkeley National Laboratory, Berkeley, USA
}%
\affiliation{%
 Dept. of Molecular and Cell Biology, Univeristy of California, Berkeley, USA
}%

\author{Holger M{\"u}ller}%
\email{hm@berkeley.edu}
\affiliation{%
 Dept. of Physics, University of California, Berkeley, USA
}%
\affiliation{%
Lawrence Berkeley National Laboratory, Berkeley, USA
}%

\date{\today}

\begin{abstract}
The secular dynamics of a non-relativistic charged particle in an electromagnetic wave can be described by the ponderomotive potential. Although ponderomotive electron-laser interactions at relativistic velocities are important for emerging technologies from laser-based particle accelerators to laser-enhanced electron microscopy, the effects of special relativity on the interaction have only been studied theoretically. Here, we use a transmission electron microscope to measure the position-dependent phase shift imparted to a relativistic electron wave function when it traverses a standing laser wave. The kinetic energy of the electrons is varied between $80\,\mathrm{keV}$ and $300\,\mathrm{keV}$, and the laser standing wave has a continuous-wave intensity of $175\,\mathrm{GW/cm}^2$. In contrast to the non-relativistic case, we demonstrate that the phase shift depends on both the electron velocity and the wave polarization, confirming the predictions of a quasiclassical theory of the interaction. Remarkably, if the electron's speed is greater than $1/\sqrt{2}$ of the speed of light, the phase shift at the electric field nodes of the wave can exceed that at the antinodes. In this case there exists a polarization such that the phase shifts at the nodes and antinodes are equal, and the electron does not experience Kapitza-Dirac diffraction. Our results thus provide new capabilities for coherent electron beam manipulation.
\end{abstract}

\maketitle


The motion of a non-relativistic charged particle in an electromagnetic (EM) wave can be described on time scales longer than the wave period by the ponderomotive potential \cite{kibble_refraction_1966}, an effective potential proportional to the time-averaged square of the electric field and independent of the EM wave polarization or particle velocity. The ponderomotive potential plays an important role in a variety of phenomena including the Kapitza-Dirac effect \cite{kapitza_reflection_1933,Freimund2001}, high-harmonic generation \cite{PhysRevA.49.2117}, laser-driven particle acceleration \cite{RevModPhys.81.1229}, free electron laser seeding \cite{RevModPhys.86.897}, electron pulse train generation \cite{PhysRevLett.120.103203,Kozak2017}, and laser-controlled electron interferometry \cite{muller_design_2010,Schwartz:17,schwartz_laser_2018}. 

A significant theoretical effort has been dedicated to generalizing the ponderomotive potential for particles with relativistic initial velocities \cite{PhysRevA.83.063810,PhysRevLett.95.053601,PhysRevA.72.043401,startsev_multiple_1997,PhysRevLett.75.4622,grebogi_relativistic_1984}. In this case, the interaction depends on both the particle velocity and EM wave polarization. While non-relativistic particles are always pushed away from the high electric field amplitude regions of the wave, relativistic particles can be deflected towards them, in an effect called relativistic reversal \cite{PhysRevA.72.043401}. This effect enables polarization-based control of the coherent manipulation of relativistic electron beams using laser light, with applications including rapidly-switchable electron beamsplitters, Kapitza-Dirac diffraction-free phase shifters, and ponderomotive free-electron laser wigglers \cite{Balcou2010,PhysRevA.83.063810}.

Here, we experimentally study the interaction of a relativistic electron with a standing laser wave. We first formulate a quasiclassical theory of the interaction which allows us to calculate the phase shift imparted to an electron wavepacket as it traverses the laser wave, from which modifications to the ponderomotive potential can be derived. Then, using the relativistic electron beam of a transmission electron microscope (TEM) and the standing laser wave of a Fabry-P{\'e}rot optical cavity, we image the phase shift imparted to the electron beam and observe velocity- and polarization-dependent relativistic effects including relativistic reversal. 

\begin{figure*} 
\includegraphics[width=17.2cm]{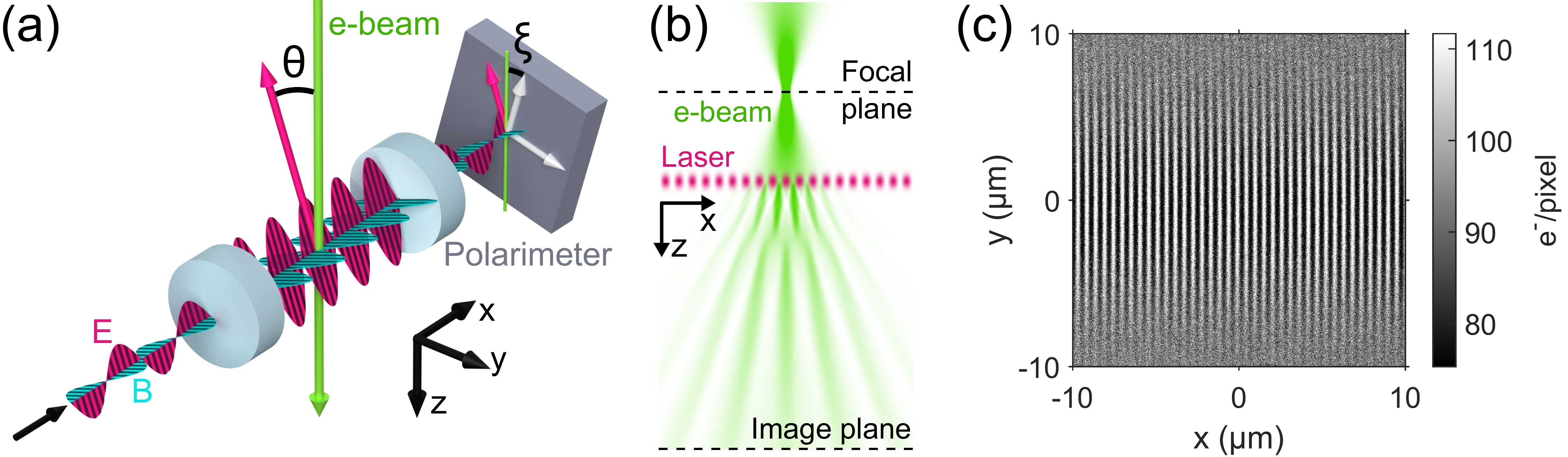}
\caption{\label{fig:schematic} \textbf{a) Schematic.} An electron beam intersects a standing laser wave (electric field shown in magenta, magnetic field shown in cyan) formed between the two mirrors (blue cylinders) of a Fabry-P{\'e}rot optical cavity. The dimensions of the cavity are not shown to scale. The polarization axis of the standing wave (magenta arrows) makes an angle $\theta$ with the electron beam axis. The polarimeter is tilted by an angle $\xi$ relative to the electron beam axis. \textbf{b) Phase modulation detection scheme.} The electron beam crosses the standing laser wave after passing through a focus. The interaction with the standing wave imprints a spatial phase modulation on the electron beam which is converted to intensity modulation as the electron beam propagates to the image plane. \textbf{c) Ronchigram of the standing laser wave.} The direct electron detection camera records the number of electrons landing on each of its pixels. In the image plane, the standing wave structure of the phase modulation manifests as a series of bright and dark fringes.}
\end{figure*}

To calculate the phase shift imparted to a relativistic charged particle by an EM wave of arbitrary spatial and temporal configuration, we use the quasiclassical approximation, which assumes that the shortest wavelength present in the EM wave $\lambda_L$ is much larger than the electron wavelength $\lambda_e$. This condition is satisfied in most experimentally relevant situations. The phase shift is then given by the action along the classical trajectory divided by the reduced Planck constant $\hbar$. We perform this calculation up to second order in the electric field strength, which in the quantum picture corresponds to stimulated Compton scattering (SCS) \cite{kapitza_reflection_1933,bartell_observation_1965}. This approximation is valid over a wide range of EM wave intensities, from an onset where SCS overcomes spontaneous Compton scattering \cite{kapitza_reflection_1933}, up to the ``relativistic" wave intensity \cite{mourou_optics_2006,bucksbaum_high-intensity_1988} where the particle is accelerated to a relativistic velocity within a single cycle of the wave.

Since the action is a Lorentz scalar, it is convenient to perform the calculations in the reference frame co-moving with the initial (unperturbed) velocity of the electron, $\mathbf{v}_0 = c \beta \hat{\mathbf{z}}$, where the motion remains non-relativistic at all times. $\beta$ is the electron's speed in units of the speed of light $c$. Variables in the co-moving frame are denoted by an apostrophe. The phase shift can then be written as
\begin{equation} \label{eqn:phaseintegral}
    \phi = \frac{1}{\hbar}\int dt' \, \left( \frac{1}{2}m \mathbf{v}'^2\left(t'\right) - e \mathbf{A}'\left(\mathbf{r}'\left(t'\right),t'\right) \cdot \mathbf{v}'\left(t'\right) \right) \,,
\end{equation}
where $m$ is the electron mass, $e$ is the elementary charge, and $\mathbf{A}'\left(\mathbf{r}'\left(t'\right),t'\right)$ is the vector potential (in the Coulomb gauge) evaluated at the electron's position $\mathbf{r}'\left(t'\right)$. We evaluate this expression perturbatively to the leading (second) order in the field strength parameter $e \left|\mathbf{A}'\right|/mc$. The first order contribution to the phase shift vanishes due to energy-momentum conservation. In the co-moving frame, the electron is initially at rest at position $\mathbf{r}'_0$. The electric field of the EM wave accelerates it to a velocity $\mathbf{v}'_1\left(t'\right)$ which, to first order in $e \left|\mathbf{A}'\right|/mc$, is given by $\mathbf{v}'_1\left(t'\right) = e \mathbf{A}'\left(\mathbf{r}'_0,t'\right)/m$. Using this expression in Eq. \eqref{eqn:phaseintegral}, the phase can then be expressed (to second order in $e \left|\mathbf{A}'\right|/mc$) in terms of only $\mathbf{A}'$:
\begin{equation}
    \phi = - \frac{1}{\hbar} \int dt' \, \frac{e^2}{2m} \mathbf{A}'^2\left(\mathbf{r}'_0,t'\right)  \,.  \label{eqn:phasecomoving}
\end{equation}
To express the phase shift in terms of the laboratory frame Coulomb gauge vector potential, we perform a Lorentz transformation and then restore the Coulomb gauge by a gauge transformation (see supplementary materials). The resulting expression is
\begin{align}
    \phi &= - \frac{1}{\hbar}  \int dt \, \frac{e^2}{2 m \gamma}  \left[  \left( \mathbf{A} \left(\mathbf{r}_0\left(t\right),t\right) - \nabla G \left(\mathbf{r}_0\left(t\right),t\right)  \right)^2 \right. \nonumber\\
    &- \beta^2 \left. \left( A_z \left(\mathbf{r}_0\left(t\right),t\right) - \nabla_z G \left(\mathbf{r}_0\left(t\right),t\right)  \right)^2    \right]  \,, \label{eqn:phaselab}
\end{align}
where $\mathbf{r}_0\left(t\right)$ is the unperturbed electron trajectory in the laboratory frame, $\gamma = \left(1-\beta^2\right)^{-1/2}$, and 
\begin{align}
    G\left(\mathbf{x},t\right) &= c \beta \int_{-\infty}^{t} dT\, A_z \left( \mathbf{x} - c\beta \left(t - T\right) \hat{\mathbf{z}} , T \right)
\end{align}
is the gauge function, and $\mathbf{x}=\left(x,y,z\right)$. Using the slowly-varying envelope approximation, which assumes that the amplitude of the EM wave varies slowly along the electron trajectory relative to its oscillation period, we can time-average the integrand of Eq. \eqref{eqn:phaselab} over one cycle of the field, leaving an effective potential. To zeroth order in $\beta$, this potential is simply the ponderomotive potential $U_p\left(\mathbf{x}\right) = \frac{e^2}{2m} \left< \mathbf{A}^2 \left(\mathbf{x},t\right) \right>$, where the angle brackets denote a time-average over one oscillation period of the field. We note, however, that Eq. \eqref{eqn:phaselab} remains valid in general for the SCS phase shift, even if the electromagnetic field does not have a slowly-varying envelope in time and space.

When the electron is relativistic, the $\beta$-dependent terms in Eq. \eqref{eqn:phaselab} cannot be neglected. In particular, the $\nabla G$ terms become relevant if the amplitude of the EM wave varies substantially over distances comparable to its wavelength, such as in a standing wave. In the case of a monochromatic standing wave with its wavevector parallel to the $x$-axis and polarization specified by angle $\theta$ and ellipticity parameter $\epsilon$, the Coulomb gauge vector potential can be written as
\begin{align}
    \mathbf{A}\left(\mathbf{x},t\right) &= A_0\left(y,z\right) \cos\left(2 \pi x /\lambda_L\right) \nonumber\\
    &\times \left[ \cos\left(\theta\right) \cos\left(\omega t\right) \hat{\mathbf{z}} + \sin\left(\theta\right) \cos\left(\omega t - \epsilon \right) \hat{\mathbf{y}}\right] \,, \label{eqn:vectorpotential}
\end{align}
where $A_0\left(y,z\right)$ is the wave's amplitude envelope and $\omega = 2\pi c / \lambda_L$ is its angular frequency. If the slowly-varying envelope approximation is satisfied, time-averaging the integrand of Eq. \eqref{eqn:phaselab} results in the relativistic effective potential
\begin{align}
    U_r \left(\mathbf{x}\right) &= \frac{e^2 A_0^2\left(y,z\right)}{4 m \gamma}  \frac{1}{2} \left[ 1 + \rho\left(\theta,\beta\right) \cos\left(4\pi x /\lambda_L\right) \right] \,, \label{eqn:pprel}
\end{align}
where
\begin{align}
    \rho\left(\theta,\beta\right) &= 1 - 2 \beta^2 \cos^2\left(\theta\right) \label{eqn:rho}
\end{align}
describes the relative depth of the standing wave structure of the potential. An electron beam passing through such an EM wave will acquire a spatial phase modulation
\begin{align}
    \phi\left(x,y\right) &= - \phi_0\left(y\right)  \frac{1}{2} \left[ 1 + \rho\left(\theta,\beta\right) \cos\left(4\pi x /\lambda_L\right) \right] \,, \label{eqn:phaserel} \\
    \phi_0\left(y\right) &\equiv  \frac{1}{\hbar} \int dz\, \frac{e^2 A_0^2\left(y,z\right)}{4 m c \beta \gamma} \,, \label{eqn:phi0y}
\end{align}
where $\phi_0\left(0\right)  \rho\left(\theta,\beta\right)$ is the depth of the phase modulation along the wave axis.

Equation \eqref{eqn:pprel} and Eq. \eqref{eqn:rho} show that the relativistic interaction is strongly dependent on both the electron speed $\beta$ and EM wave polarization angle $\theta$, though not on the ellipticity parameter $\epsilon$. Importantly, if $\beta \geq 1/\sqrt{2}$, there exists a polarization angle $\theta_r$, referred to as the relativistic reversal angle \cite{PhysRevA.72.043401}, such that $\rho\left(\theta_r,\beta\right)=0$. At this angle, the standing wave structure of the phase shift disappears entirely, and therefore no Kapitza-Dirac diffraction occurs. 

The relativistic interaction also modifies the laser-induced group delay of the electron wave function. The group delay, equivalent to the retardation of a classical particle and defined as $\tau = \hbar \frac{d \phi}{d K}$, can be calculated from the energy dependence of the electron phase shift:
\begin{align}
\tau\left(x,y\right) &= \frac{ \hbar}{ m c^2} \frac{\phi_0\left(y\right)}{\beta^2 \gamma}  \frac{1}{2}\left[ 1+ \varrho\left(\theta,\beta\right) \cos\left(4 \pi x / \lambda_L\right)  \right] \,, \label{eqn:gd} \\
\varrho\left(\theta,\beta\right) &\equiv 1+ 2\beta^2 \left(1-2 \beta^2\right) \cos^2\left(\theta\right) \,. \label{eqn:gdpol}
\end{align}
In particular, at $\theta=\theta_r$, when the standing-wave structure in the potential of Eq. \eqref{eqn:pprel} vanishes, the standing wave structure is still present in the spatial profile of the group delay. Furthermore, when $0<\beta<1/\sqrt{2}$, the group delay is negative for portions of the standing wave around the electric field nodes. This negative group delay corresponds to an attractive potential, in contrast to the non-relativistic ponderomotive potential which is always repulsive.

A schematic of the experiment is shown in Fig. \ref{fig:schematic}(a). The electron beam of a TEM (Thermo Fisher Scientific Titan) passes through a standing laser wave, where the axis of the standing wave $\hat{x}$ is perpendicular to the propagation direction of the electron beam $\hat{z}$. The interaction imprints a spatial phase modulation on the electron wave function, as described by Eq. \eqref{eqn:phaserel}. The electron beam then propagates away from the interaction region before it is imaged using a direct electron detection camera (Gatan K2) \cite{Ruskin2013}. The electron beam is brought to a focus before it crosses the standing laser wave such that a point-projection image, known as a ``Ronchigram," is formed on the camera \cite{spence2013high,Lupini2011}. As illustrated in Fig. \ref{fig:schematic}(b), paraxial propagation of the electron beam from the interaction region to the camera partially converts the phase modulation of the electron wave function to amplitude modulation, allowing the phase modulation to be imaged. The electron's kinetic energy $K = \left(\gamma-1\right) m c^2$ can be adjusted between $80\,\mathrm{keV}$ and $300\,\mathrm{keV}$ by changing the TEM's accelerating voltage.

The standing laser wave is formed inside of a Fabry-P{\'e}rot optical cavity which serves to amplify and focus a continuous-wave laser beam with a wavelength of $\lambda_L = 1064\,\mathrm{nm}$ \cite{muller_design_2010,Schwartz:17,schwartz_laser_2018}. The fundamental mode of the cavity has a Gaussian profile such that at its waist where it intersects the electron beam, Eq. \eqref{eqn:phi0y} gives
\begin{align}
    \phi_0\left(y\right) &= e^{-2 \frac{y^2}{w_0^2}}  \sqrt{\frac{8}{\pi^3}} \frac{\alpha}{\beta \gamma} \frac{P \lambda_L^2}{m c^3 w_0} \,, \label{eqn:phi0}
\end{align} 
where $w_0$ is the $1/e^2$ radius of the mode, $\alpha$ is the fine-structure constant, and $P$ is the optical power circulating in the cavity. Since a relativistic electron spends little time interacting with the EM wave, the laser intensity must be high in order for the electron wave function to accumulate appreciable phase. We achieve phase shifts on the order of $1\,\mathrm{rad}$ with a circulating power of $44\,\mathrm{kW}$ focused to a $w_0 = 8\,\mu\mathrm{m}$ focus, corresponding to a peak standing wave intensity of $175\,\mathrm{GW}/\mathrm{cm}^2$. To our knowledge, this is the highest continuous-wave laser intensity ever achieved.

A half-wave plate placed at the input of the optical cavity is used to control the linear polarization angle of the light entering the cavity. A portion of the light transmitted through the cavity is sent to a polarimeter which measures the optical power in the two orthogonal polarization components relative to the polarimeter axis. The polarimeter employs polarizing beamsplitter cubes to separate the orthogonal polarization components, and calibrated photodiodes to measure the optical power of each component (see supplementary materials). The absolute value and sign of the polarization angle $\theta$ are determined from the polarimeter reading and orientation of the half-wave plate, respectively. The polarization at the polarimeter is assumed to be the same as that inside the cavity, as the cavity was measured to not appreciably change the polarization between its input and output (see supplementary materials).

\begin{figure*} 
\includegraphics[width=17.2cm]{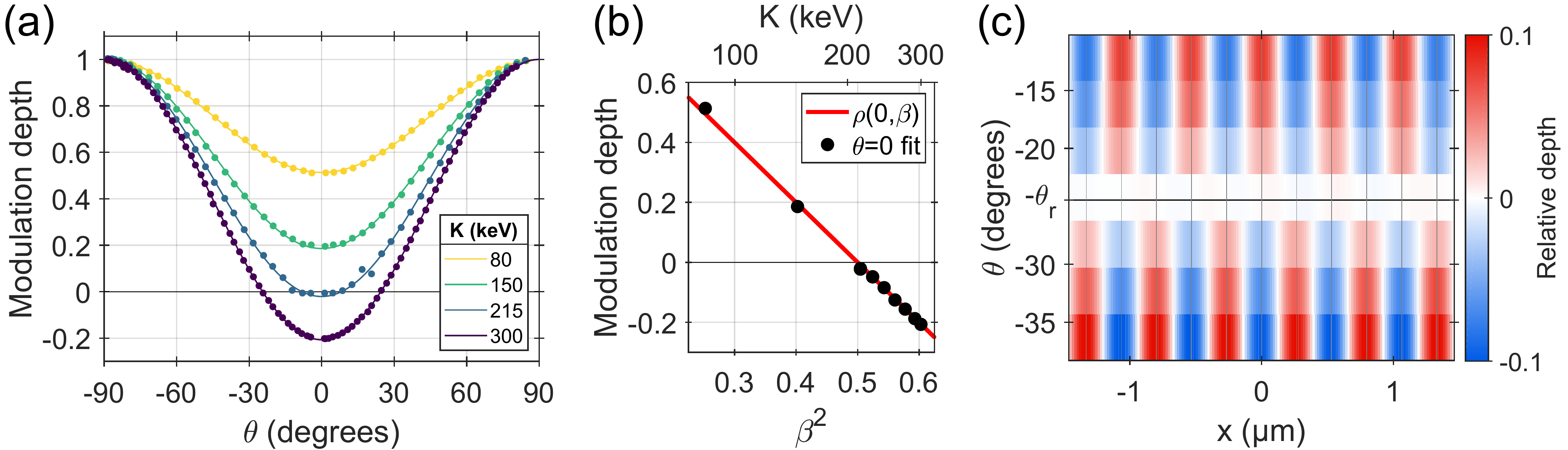}
\caption{\label{fig:data1} \textbf{a) Relative phase modulation depth as a function of polarization angle.} Measurements of the relative phase modulation depth (dots) are plotted along with fitted theory curves (lines) as a function of the laser wave polarization angle $\theta$ for several values of the nominal electron energy $K$ (equivalently, electron speed $\beta$). The fitted theory (solid lines) is described by Eq. \eqref{eqn:rho}. \textbf{b) Relative phase modulation depth as a function of electron speed.} Fit values for the relative phase modulation depth at $\theta = 0$ are shown as a function of $\beta^2$ for each of the nine electron energies examined (black dots). The theoretical dependence on $\beta^2$ predicted by Eq. \eqref{eqn:rho} is shown in red. \textbf{c) Relativistic reversal.} Ronchigram standing wave fringes are shown as a function of polarization angle $\theta$ and position along the laser beam axis $x$ for the $K= 300\, \mathrm{keV}$ data set used in panel (a). The bright (red) and dark (blue) fringes reverse positions around the relativistic reversal angle $\theta_r$ (horizontal black line). The vertical lines are a guide-to-eye.}
\end{figure*}

Ronchigrams were collected at electron beam energies of $K= 80$, 150, 215, 230, 245, 260, 275, 290, and $300 \,\mathrm{keV}$. At each electron energy the rotation angle of the half-wave plate was incremented between $10\,\mathrm{s}$ electron camera exposures. The half-wave plate was rotated through $90^\circ$ in one direction and then rotated back to the original position, thereby rotating the polarization angle from $\theta\approx-90^\circ$ to $\theta\approx+90^\circ$ and back again. Rotation of the polarization angle through a full $180^\circ$ allowed for the determination of any misalignment between the polarimeter axis and the electron beam axis (angle $\xi$ shown in Fig. \ref{fig:schematic}(a)).

Figure \ref{fig:schematic}(c) shows a typical unprocessed Ronchigram. Both the standing wave structure and the transverse Gaussian profile of the cavity mode are clearly evident. Each Ronchigram was fit in Fourier space using the phase modulation depth of the standing wave as a fit parameter (see supplementary materials). To correct for small variations in the laser wave parameters during the experiment, the phase modulation depth was normalized by the optical power at the polarimeter (proportional to the circulating power $P$) and the mode waist $w_0$, which were both measured at the time the Ronchigram was taken. The mode waist was determined from a measurement of the cavity's transverse mode frequency spacing (see supplementary materials). The fractional change in $\lambda_L$ during a typical experiment was measured to be small enough ($\sim 10^{-6}$) that it was assumed to be constant for the purpose of normalization. 

Each set of normalized modulation depth versus polarization angle data was fit to Eq. \eqref{eqn:rho}, with an angle-independent normalization constant, a polarization angle axis offset, and the electron speed $\beta$ as fit parameters. The electron speed was used as a fit parameter because the nominal electron energies $K$ are only accurate to approximately $\pm 1 \%$ (per the TEM manufacturer's specifications). The polarization angle axis offset accounts for the polarimeter misalignment angle $\xi$, as well as any linear polarization rotation induced by optics between the cavity output and polarimeter (see supplementary materials).

This data is presented in Fig. \ref{fig:data1}(a); for clarity, only the $K = 80$, 150, 215, and $300 \, \mathrm{keV}$ data sets are shown. The remaining data sets are shown in the supplementary materials. The relative phase modulation depth exhibits a dependence on the polarization angle $\theta$ that is well-modeled by Eq. \eqref{eqn:rho}; the root-mean-squared difference between the fit and data across all data sets is $7.1\times10^{-3}$. To show the relative phase modulation depth's $\beta$-dependence, the fit values at $\theta=0$ are plotted as a function of the nominal values of $\beta^2$ for all data sets in Fig. \ref{fig:data1}(b), where they are compared with the linear dependence expected from Eq. \eqref{eqn:rho}. Again, there is a good correspondence between the measured values and theoretical model.

Imaging of the spatial phase modulation profile allows the relativistic reversal effect to be directly observed in the Ronchigrams. As the polarization angle is rotated through the relativistic reversal angle, the standing wave structure in the Ronchigram diminishes in amplitude until it disappears entirely and then re-emerges with the opposite sign. This is demonstrated in Fig. \ref{fig:data1}(c), using Ronchigrams from the $K = 300\,\mathrm{keV}$ data set shown in Fig. \ref{fig:data1}(a). A temporally linear drift in the fringe position of $0.4\, \mathrm{nm}/\mathrm{s}$ due to thermal expansion of the cavity support structure has been removed from the displayed data (see supplementary materials). The change in fringe position around $\theta_r$ was used to infer the sign of the relative phase modulation depth for the $K \geq 215\,\mathrm{keV}$ data sets in Fig. \ref{fig:data1}(a) and Fig. \ref{fig:data1}(b).

Our results show that SCS of relativistic particles exhibits a strong dependence on the electron velocity and the EM wave polarization, and that this dependence is well-described by a quasiclassical theory of the interaction. The most striking feature of the polarization dependence is that the standing wave structure of the phase shift reverses its sign at a particular polarization angle when $\beta\geq1/\sqrt{2}$. Therefore, this experiment can also be understood as an observation of the relativistic reversal of the amplitude of Kapitza-Dirac diffraction \cite{kapitza_reflection_1933,Freimund2001}.

The dependence of the relativistic effective potential on the polarization of the EM wave provides an avenue for dynamical optical control of relativistic electron beams. When the standing wave structure of the phase modulation is eliminated, the electron wave function does not diffract from the EM wave. Therefore, varying the polarization of the standing wave could be used to make a rapidly-switchable electron beamsplitter, or implement electron pulse slicing \cite{Kanters2011}. The same effect could be used to temporally phase modulate an electron beam focused through a single antinode of the standing wave. Additionally, when the standing EM wave is used as a phase plate for phase contrast electron microscopy, operation at the relativistic reversal angle eliminates the presence of ``ghost" images due to diffraction (see supplementary materials) \cite{Schwartz:17,schwartz_laser_2018}. Polarization-based control of the relativistic effective potential thus adds a much-needed capability to the presently sparse toolkit for coherent electron beam manipulation.

\begin{acknowledgments}
We thank P. Ahmadi, B. Buijsse, W. T. Carlisle, R. Danev, P. Dona, S. Goobie, H. Green, P. Grob, F. Littlefield, G. W. Long, J. Lopez, E. Nogales, and B. W. Reed for their contributions to this work. This work was supported by the US National Institutes of Health grant 5 R01 GM126011-02, and Bakar Fellows Program. JJA is supported by the National Science Foundation Graduate Research Fellowship Program Grant No. DGE 1752814. SLC is supported by the Howard Hughes Medical Institute Hanna H. Gray Fellows Program Grant No. GT11085.
\end{acknowledgments}

\nocite{ucb.b1269885920070101}

\providecommand{\noopsort}[1]{}\providecommand{\singleletter}[1]{#1}%
\begin{thebibliography}{26}%
\makeatletter
\providecommand \@ifxundefined [1]{%
 \@ifx{#1\undefined}
}%
\providecommand \@ifnum [1]{%
 \ifnum #1\expandafter \@firstoftwo
 \else \expandafter \@secondoftwo
 \fi
}%
\providecommand \@ifx [1]{%
 \ifx #1\expandafter \@firstoftwo
 \else \expandafter \@secondoftwo
 \fi
}%
\providecommand \natexlab [1]{#1}%
\providecommand \enquote  [1]{``#1''}%
\providecommand \bibnamefont  [1]{#1}%
\providecommand \bibfnamefont [1]{#1}%
\providecommand \citenamefont [1]{#1}%
\providecommand \href@noop [0]{\@secondoftwo}%
\providecommand \href [0]{\begingroup \@sanitize@url \@href}%
\providecommand \@href[1]{\@@startlink{#1}\@@href}%
\providecommand \@@href[1]{\endgroup#1\@@endlink}%
\providecommand \@sanitize@url [0]{\catcode `\\12\catcode `\$12\catcode
  `\&12\catcode `\#12\catcode `\^12\catcode `\_12\catcode `\%12\relax}%
\providecommand \@@startlink[1]{}%
\providecommand \@@endlink[0]{}%
\providecommand \url  [0]{\begingroup\@sanitize@url \@url }%
\providecommand \@url [1]{\endgroup\@href {#1}{\urlprefix }}%
\providecommand \urlprefix  [0]{URL }%
\providecommand \Eprint [0]{\href }%
\providecommand \doibase [0]{https://doi.org/}%
\providecommand \selectlanguage [0]{\@gobble}%
\providecommand \bibinfo  [0]{\@secondoftwo}%
\providecommand \bibfield  [0]{\@secondoftwo}%
\providecommand \translation [1]{[#1]}%
\providecommand \BibitemOpen [0]{}%
\providecommand \bibitemStop [0]{}%
\providecommand \bibitemNoStop [0]{.\EOS\space}%
\providecommand \EOS [0]{\spacefactor3000\relax}%
\providecommand \BibitemShut  [1]{\csname bibitem#1\endcsname}%
\let\auto@bib@innerbib\@empty
\bibitem [{\citenamefont {Kibble}(1966)}]{kibble_refraction_1966}%
  \BibitemOpen
  \bibfield  {author} {\bibinfo {author} {\bibfnamefont {T.~W.~B.}\
  \bibnamefont {Kibble}},\ }\bibfield  {title} {\bibinfo {title} {Refraction of
  {Electron} {Beams} by {Intense} {Electromagnetic} {Waves}},\ }\href
  {https://doi.org/10.1103/PhysRevLett.16.1054} {\bibfield  {journal} {\bibinfo
   {journal} {Physical Review Letters}\ }\textbf {\bibinfo {volume} {16}},\
  \bibinfo {pages} {1054} (\bibinfo {year} {1966})}\BibitemShut {NoStop}%
\bibitem [{\citenamefont {Kapitza}\ and\ \citenamefont
  {Dirac}(1933)}]{kapitza_reflection_1933}%
  \BibitemOpen
  \bibfield  {author} {\bibinfo {author} {\bibfnamefont {P.~L.}\ \bibnamefont
  {Kapitza}}\ and\ \bibinfo {author} {\bibfnamefont {P.~A.~M.}\ \bibnamefont
  {Dirac}},\ }\bibfield  {title} {\bibinfo {title} {The reflection of electrons
  from standing light waves},\ }\href
  {https://doi.org/10.1017/S0305004100011105} {\bibfield  {journal} {\bibinfo
  {journal} {Mathematical Proceedings of the Cambridge Philosophical Society}\
  }\textbf {\bibinfo {volume} {29}},\ \bibinfo {pages} {297} (\bibinfo {year}
  {1933})}\BibitemShut {NoStop}%
\bibitem [{\citenamefont {Freimund}\ \emph {et~al.}(2001)\citenamefont
  {Freimund}, \citenamefont {Aflatooni},\ and\ \citenamefont
  {Batelaan}}]{Freimund2001}%
  \BibitemOpen
  \bibfield  {author} {\bibinfo {author} {\bibfnamefont {D.~L.}\ \bibnamefont
  {Freimund}}, \bibinfo {author} {\bibfnamefont {K.}~\bibnamefont
  {Aflatooni}},\ and\ \bibinfo {author} {\bibfnamefont {H.}~\bibnamefont
  {Batelaan}},\ }\bibfield  {title} {\bibinfo {title} {Observation of the
  kapitza-dirac effect},\ }\href {https://doi.org/10.1038/35093065} {\bibfield
  {journal} {\bibinfo  {journal} {Nature}\ }\textbf {\bibinfo {volume} {413}},\
  \bibinfo {pages} {142} (\bibinfo {year} {2001})}\BibitemShut {NoStop}%
\bibitem [{\citenamefont {Lewenstein}\ \emph {et~al.}(1994)\citenamefont
  {Lewenstein}, \citenamefont {Balcou}, \citenamefont {Ivanov}, \citenamefont
  {L'Huillier},\ and\ \citenamefont {Corkum}}]{PhysRevA.49.2117}%
  \BibitemOpen
  \bibfield  {author} {\bibinfo {author} {\bibfnamefont {M.}~\bibnamefont
  {Lewenstein}}, \bibinfo {author} {\bibfnamefont {P.}~\bibnamefont {Balcou}},
  \bibinfo {author} {\bibfnamefont {M.~Y.}\ \bibnamefont {Ivanov}}, \bibinfo
  {author} {\bibfnamefont {A.}~\bibnamefont {L'Huillier}},\ and\ \bibinfo
  {author} {\bibfnamefont {P.~B.}\ \bibnamefont {Corkum}},\ }\bibfield  {title}
  {\bibinfo {title} {Theory of high-harmonic generation by low-frequency laser
  fields},\ }\href {https://doi.org/10.1103/PhysRevA.49.2117} {\bibfield
  {journal} {\bibinfo  {journal} {Phys. Rev. A}\ }\textbf {\bibinfo {volume}
  {49}},\ \bibinfo {pages} {2117} (\bibinfo {year} {1994})}\BibitemShut
  {NoStop}%
\bibitem [{\citenamefont {Esarey}\ \emph {et~al.}(2009)\citenamefont {Esarey},
  \citenamefont {Schroeder},\ and\ \citenamefont
  {Leemans}}]{RevModPhys.81.1229}%
  \BibitemOpen
  \bibfield  {author} {\bibinfo {author} {\bibfnamefont {E.}~\bibnamefont
  {Esarey}}, \bibinfo {author} {\bibfnamefont {C.~B.}\ \bibnamefont
  {Schroeder}},\ and\ \bibinfo {author} {\bibfnamefont {W.~P.}\ \bibnamefont
  {Leemans}},\ }\bibfield  {title} {\bibinfo {title} {Physics of laser-driven
  plasma-based electron accelerators},\ }\href
  {https://doi.org/10.1103/RevModPhys.81.1229} {\bibfield  {journal} {\bibinfo
  {journal} {Rev. Mod. Phys.}\ }\textbf {\bibinfo {volume} {81}},\ \bibinfo
  {pages} {1229} (\bibinfo {year} {2009})}\BibitemShut {NoStop}%
\bibitem [{\citenamefont {Hemsing}\ \emph {et~al.}(2014)\citenamefont
  {Hemsing}, \citenamefont {Stupakov}, \citenamefont {Xiang},\ and\
  \citenamefont {Zholents}}]{RevModPhys.86.897}%
  \BibitemOpen
  \bibfield  {author} {\bibinfo {author} {\bibfnamefont {E.}~\bibnamefont
  {Hemsing}}, \bibinfo {author} {\bibfnamefont {G.}~\bibnamefont {Stupakov}},
  \bibinfo {author} {\bibfnamefont {D.}~\bibnamefont {Xiang}},\ and\ \bibinfo
  {author} {\bibfnamefont {A.}~\bibnamefont {Zholents}},\ }\bibfield  {title}
  {\bibinfo {title} {Beam by design: Laser manipulation of electrons in modern
  accelerators},\ }\href {https://doi.org/10.1103/RevModPhys.86.897} {\bibfield
   {journal} {\bibinfo  {journal} {Rev. Mod. Phys.}\ }\textbf {\bibinfo
  {volume} {86}},\ \bibinfo {pages} {897} (\bibinfo {year} {2014})}\BibitemShut
  {NoStop}%
\bibitem [{\citenamefont {Koz{\'a}k}\ \emph {et~al.}(2018)\citenamefont
  {Koz{\'a}k}, \citenamefont {Sch{\"o}nenberger},\ and\ \citenamefont
  {Hommelhoff}}]{PhysRevLett.120.103203}%
  \BibitemOpen
  \bibfield  {author} {\bibinfo {author} {\bibfnamefont {M.}~\bibnamefont
  {Koz{\'a}k}}, \bibinfo {author} {\bibfnamefont {N.}~\bibnamefont
  {Sch{\"o}nenberger}},\ and\ \bibinfo {author} {\bibfnamefont
  {P.}~\bibnamefont {Hommelhoff}},\ }\bibfield  {title} {\bibinfo {title}
  {Ponderomotive generation and detection of attosecond free-electron pulse
  trains},\ }\href {https://doi.org/10.1103/PhysRevLett.120.103203} {\bibfield
  {journal} {\bibinfo  {journal} {Phys. Rev. Lett.}\ }\textbf {\bibinfo
  {volume} {120}},\ \bibinfo {pages} {103203} (\bibinfo {year}
  {2018})}\BibitemShut {NoStop}%
\bibitem [{\citenamefont {Koz{\'a}k}\ \emph {et~al.}(2017)\citenamefont
  {Koz{\'a}k}, \citenamefont {Eckstein}, \citenamefont {Sch{\"o}nenberger},\
  and\ \citenamefont {Hommelhoff}}]{Kozak2017}%
  \BibitemOpen
  \bibfield  {author} {\bibinfo {author} {\bibfnamefont {M.}~\bibnamefont
  {Koz{\'a}k}}, \bibinfo {author} {\bibfnamefont {T.}~\bibnamefont {Eckstein}},
  \bibinfo {author} {\bibfnamefont {N.}~\bibnamefont {Sch{\"o}nenberger}},\
  and\ \bibinfo {author} {\bibfnamefont {P.}~\bibnamefont {Hommelhoff}},\
  }\bibfield  {title} {\bibinfo {title} {Inelastic ponderomotive scattering of
  electrons at a high-intensity optical travelling wave in vacuum},\ }\href
  {https://doi.org/10.1038/nphys4282} {\bibfield  {journal} {\bibinfo
  {journal} {Nature Physics}\ }\textbf {\bibinfo {volume} {14}},\ \bibinfo
  {pages} {121} (\bibinfo {year} {2017})}\BibitemShut {NoStop}%
\bibitem [{\citenamefont {M{\"u}ller}\ \emph {et~al.}(2010)\citenamefont
  {M{\"u}ller}, \citenamefont {Jin}, \citenamefont {Danev}, \citenamefont
  {Spence}, \citenamefont {Padmore},\ and\ \citenamefont
  {Glaeser}}]{muller_design_2010}%
  \BibitemOpen
  \bibfield  {author} {\bibinfo {author} {\bibfnamefont {H.}~\bibnamefont
  {M{\"u}ller}}, \bibinfo {author} {\bibfnamefont {J.}~\bibnamefont {Jin}},
  \bibinfo {author} {\bibfnamefont {R.}~\bibnamefont {Danev}}, \bibinfo
  {author} {\bibfnamefont {J.}~\bibnamefont {Spence}}, \bibinfo {author}
  {\bibfnamefont {H.}~\bibnamefont {Padmore}},\ and\ \bibinfo {author}
  {\bibfnamefont {R.~M.}\ \bibnamefont {Glaeser}},\ }\bibfield  {title}
  {\bibinfo {title} {Design of an electron microscope phase plate using a
  focused continuous-wave laser},\ }\href
  {https://doi.org/10.1088/1367-2630/12/7/073011} {\bibfield  {journal}
  {\bibinfo  {journal} {New Journal of Physics}\ }\textbf {\bibinfo {volume}
  {12}},\ \bibinfo {pages} {073011} (\bibinfo {year} {2010})}\BibitemShut
  {NoStop}%
\bibitem [{\citenamefont {Schwartz}\ \emph {et~al.}(2017)\citenamefont
  {Schwartz}, \citenamefont {Axelrod}, \citenamefont {Tuthill}, \citenamefont
  {Haslinger}, \citenamefont {Ophus}, \citenamefont {Glaeser},\ and\
  \citenamefont {M{\"u}ller}}]{Schwartz:17}%
  \BibitemOpen
  \bibfield  {author} {\bibinfo {author} {\bibfnamefont {O.}~\bibnamefont
  {Schwartz}}, \bibinfo {author} {\bibfnamefont {J.}~\bibnamefont {Axelrod}},
  \bibinfo {author} {\bibfnamefont {D.~R.}\ \bibnamefont {Tuthill}}, \bibinfo
  {author} {\bibfnamefont {P.}~\bibnamefont {Haslinger}}, \bibinfo {author}
  {\bibfnamefont {C.}~\bibnamefont {Ophus}}, \bibinfo {author} {\bibfnamefont
  {R.}~\bibnamefont {Glaeser}},\ and\ \bibinfo {author} {\bibfnamefont
  {H.}~\bibnamefont {M{\"u}ller}},\ }\bibfield  {title} {\bibinfo {title}
  {Near-concentric fabry-p{\'e}rot cavity for continuous-wave laser control of
  electron waves},\ }\href {https://doi.org/10.1364/OE.25.014453} {\bibfield
  {journal} {\bibinfo  {journal} {Opt. Express}\ }\textbf {\bibinfo {volume}
  {25}},\ \bibinfo {pages} {14453} (\bibinfo {year} {2017})}\BibitemShut
  {NoStop}%
\bibitem [{\citenamefont {Schwartz}\ \emph {et~al.}(2019)\citenamefont
  {Schwartz}, \citenamefont {Axelrod}, \citenamefont {Campbell}, \citenamefont
  {Turnbaugh}, \citenamefont {Glaeser},\ and\ \citenamefont
  {M{\"u}ller}}]{schwartz_laser_2018}%
  \BibitemOpen
  \bibfield  {author} {\bibinfo {author} {\bibfnamefont {O.}~\bibnamefont
  {Schwartz}}, \bibinfo {author} {\bibfnamefont {J.~J.}\ \bibnamefont
  {Axelrod}}, \bibinfo {author} {\bibfnamefont {S.~L.}\ \bibnamefont
  {Campbell}}, \bibinfo {author} {\bibfnamefont {C.}~\bibnamefont {Turnbaugh}},
  \bibinfo {author} {\bibfnamefont {R.~M.}\ \bibnamefont {Glaeser}},\ and\
  \bibinfo {author} {\bibfnamefont {H.}~\bibnamefont {M{\"u}ller}},\ }\bibfield
   {title} {\bibinfo {title} {Laser phase plate for transmission electron
  microscopy},\ }\href {https://doi.org/10.1038/s41592-019-0552-2} {\bibfield
  {journal} {\bibinfo  {journal} {Nature Methods}\ }\textbf {\bibinfo {volume}
  {16}},\ \bibinfo {pages} {1016} (\bibinfo {year} {2019})}\BibitemShut
  {NoStop}%
\bibitem [{\citenamefont {Smorenburg}\ \emph {et~al.}(2011)\citenamefont
  {Smorenburg}, \citenamefont {Kanters}, \citenamefont {Lassise}, \citenamefont
  {Brussaard}, \citenamefont {Kamp},\ and\ \citenamefont
  {Luiten}}]{PhysRevA.83.063810}%
  \BibitemOpen
  \bibfield  {author} {\bibinfo {author} {\bibfnamefont {P.~W.}\ \bibnamefont
  {Smorenburg}}, \bibinfo {author} {\bibfnamefont {J.~H.~M.}\ \bibnamefont
  {Kanters}}, \bibinfo {author} {\bibfnamefont {A.}~\bibnamefont {Lassise}},
  \bibinfo {author} {\bibfnamefont {G.~J.~H.}\ \bibnamefont {Brussaard}},
  \bibinfo {author} {\bibfnamefont {L.~P.~J.}\ \bibnamefont {Kamp}},\ and\
  \bibinfo {author} {\bibfnamefont {O.~J.}\ \bibnamefont {Luiten}},\ }\bibfield
   {title} {\bibinfo {title} {Polarization-dependent ponderomotive gradient
  force in a standing wave},\ }\href
  {https://doi.org/10.1103/PhysRevA.83.063810} {\bibfield  {journal} {\bibinfo
  {journal} {Phys. Rev. A}\ }\textbf {\bibinfo {volume} {83}},\ \bibinfo
  {pages} {063810} (\bibinfo {year} {2011})}\BibitemShut {NoStop}%
\bibitem [{\citenamefont {Kaplan}\ and\ \citenamefont
  {Pokrovsky}(2005)}]{PhysRevLett.95.053601}%
  \BibitemOpen
  \bibfield  {author} {\bibinfo {author} {\bibfnamefont {A.~E.}\ \bibnamefont
  {Kaplan}}\ and\ \bibinfo {author} {\bibfnamefont {A.~L.}\ \bibnamefont
  {Pokrovsky}},\ }\bibfield  {title} {\bibinfo {title} {Fully relativistic
  theory of the ponderomotive force in an ultraintense standing wave},\ }\href
  {https://doi.org/10.1103/PhysRevLett.95.053601} {\bibfield  {journal}
  {\bibinfo  {journal} {Phys. Rev. Lett.}\ }\textbf {\bibinfo {volume} {95}},\
  \bibinfo {pages} {053601} (\bibinfo {year} {2005})}\BibitemShut {NoStop}%
\bibitem [{\citenamefont {Pokrovsky}\ and\ \citenamefont
  {Kaplan}(2005)}]{PhysRevA.72.043401}%
  \BibitemOpen
  \bibfield  {author} {\bibinfo {author} {\bibfnamefont {A.~L.}\ \bibnamefont
  {Pokrovsky}}\ and\ \bibinfo {author} {\bibfnamefont {A.~E.}\ \bibnamefont
  {Kaplan}},\ }\bibfield  {title} {\bibinfo {title} {Relativistic reversal of
  the ponderomotive force in a standing laser wave},\ }\href
  {https://doi.org/10.1103/PhysRevA.72.043401} {\bibfield  {journal} {\bibinfo
  {journal} {Phys. Rev. A}\ }\textbf {\bibinfo {volume} {72}},\ \bibinfo
  {pages} {043401} (\bibinfo {year} {2005})}\BibitemShut {NoStop}%
\bibitem [{\citenamefont {Startsev}\ and\ \citenamefont
  {McKinstrie}(1997)}]{startsev_multiple_1997}%
  \BibitemOpen
  \bibfield  {author} {\bibinfo {author} {\bibfnamefont {E.~A.}\ \bibnamefont
  {Startsev}}\ and\ \bibinfo {author} {\bibfnamefont {C.~J.}\ \bibnamefont
  {McKinstrie}},\ }\bibfield  {title} {\bibinfo {title} {Multiple scale
  derivation of the relativistic ponderomotive force},\ }\href
  {https://doi.org/10.1103/PhysRevE.55.7527} {\bibfield  {journal} {\bibinfo
  {journal} {Physical Review E}\ }\textbf {\bibinfo {volume} {55}},\ \bibinfo
  {pages} {7527} (\bibinfo {year} {1997})}\BibitemShut {NoStop}%
\bibitem [{\citenamefont {Bauer}\ \emph {et~al.}(1995)\citenamefont {Bauer},
  \citenamefont {Mulser},\ and\ \citenamefont {Steeb}}]{PhysRevLett.75.4622}%
  \BibitemOpen
  \bibfield  {author} {\bibinfo {author} {\bibfnamefont {D.}~\bibnamefont
  {Bauer}}, \bibinfo {author} {\bibfnamefont {P.}~\bibnamefont {Mulser}},\ and\
  \bibinfo {author} {\bibfnamefont {W.~H.}\ \bibnamefont {Steeb}},\ }\bibfield
  {title} {\bibinfo {title} {Relativistic ponderomotive force, uphill
  acceleration, and transition to chaos},\ }\href
  {https://doi.org/10.1103/PhysRevLett.75.4622} {\bibfield  {journal} {\bibinfo
   {journal} {Phys. Rev. Lett.}\ }\textbf {\bibinfo {volume} {75}},\ \bibinfo
  {pages} {4622} (\bibinfo {year} {1995})}\BibitemShut {NoStop}%
\bibitem [{\citenamefont {Grebogi}\ and\ \citenamefont
  {Littlejohn}(1984)}]{grebogi_relativistic_1984}%
  \BibitemOpen
  \bibfield  {author} {\bibinfo {author} {\bibfnamefont {C.}~\bibnamefont
  {Grebogi}}\ and\ \bibinfo {author} {\bibfnamefont {R.~G.}\ \bibnamefont
  {Littlejohn}},\ }\bibfield  {title} {\bibinfo {title} {Relativistic
  ponderomotive {Hamiltonian}},\ }\href
  {https://aip.scitation.org/doi/abs/10.1063/1.864855} {\bibfield  {journal}
  {\bibinfo  {journal} {Physics of Fluids}\ }\textbf {\bibinfo {volume} {27}}
  (\bibinfo {year} {1984})}\BibitemShut {NoStop}%
\bibitem [{\citenamefont {Balcou}(2010)}]{Balcou2010}%
  \BibitemOpen
  \bibfield  {author} {\bibinfo {author} {\bibfnamefont {P.}~\bibnamefont
  {Balcou}},\ }\bibfield  {title} {\bibinfo {title} {Proposal for a raman x-ray
  free electron laser},\ }\href {https://doi.org/10.1140/epjd/e2010-00185-5}
  {\bibfield  {journal} {\bibinfo  {journal} {The European Physical Journal D}\
  }\textbf {\bibinfo {volume} {59}},\ \bibinfo {pages} {525} (\bibinfo {year}
  {2010})}\BibitemShut {NoStop}%
\bibitem [{\citenamefont {Bartell}\ \emph {et~al.}(1965)\citenamefont
  {Bartell}, \citenamefont {Thompson},\ and\ \citenamefont
  {Roskos}}]{bartell_observation_1965}%
  \BibitemOpen
  \bibfield  {author} {\bibinfo {author} {\bibfnamefont {L.~S.}\ \bibnamefont
  {Bartell}}, \bibinfo {author} {\bibfnamefont {H.~B.}\ \bibnamefont
  {Thompson}},\ and\ \bibinfo {author} {\bibfnamefont {R.~R.}\ \bibnamefont
  {Roskos}},\ }\bibfield  {title} {\bibinfo {title} {Observation of
  {Stimulated} {Compton} {Scattering} of {Electrons} by {Laser} {Beam}},\
  }\href {https://doi.org/10.1103/PhysRevLett.14.851} {\bibfield  {journal}
  {\bibinfo  {journal} {Physical Review Letters}\ }\textbf {\bibinfo {volume}
  {14}},\ \bibinfo {pages} {851} (\bibinfo {year} {1965})}\BibitemShut
  {NoStop}%
\bibitem [{\citenamefont {Mourou}\ \emph {et~al.}(2006)\citenamefont {Mourou},
  \citenamefont {Tajima},\ and\ \citenamefont {Bulanov}}]{mourou_optics_2006}%
  \BibitemOpen
  \bibfield  {author} {\bibinfo {author} {\bibfnamefont {G.~A.}\ \bibnamefont
  {Mourou}}, \bibinfo {author} {\bibfnamefont {T.}~\bibnamefont {Tajima}},\
  and\ \bibinfo {author} {\bibfnamefont {S.~V.}\ \bibnamefont {Bulanov}},\
  }\bibfield  {title} {\bibinfo {title} {Optics in the relativistic regime},\
  }\href {https://doi.org/10.1103/RevModPhys.78.309} {\bibfield  {journal}
  {\bibinfo  {journal} {Reviews of Modern Physics}\ }\textbf {\bibinfo {volume}
  {78}},\ \bibinfo {pages} {309} (\bibinfo {year} {2006})}\BibitemShut
  {NoStop}%
\bibitem [{\citenamefont {Bucksbaum}\ \emph {et~al.}(1988)\citenamefont
  {Bucksbaum}, \citenamefont {Schumacher},\ and\ \citenamefont
  {Bashkansky}}]{bucksbaum_high-intensity_1988}%
  \BibitemOpen
  \bibfield  {author} {\bibinfo {author} {\bibfnamefont {P.~H.}\ \bibnamefont
  {Bucksbaum}}, \bibinfo {author} {\bibfnamefont {D.~W.}\ \bibnamefont
  {Schumacher}},\ and\ \bibinfo {author} {\bibfnamefont {M.}~\bibnamefont
  {Bashkansky}},\ }\bibfield  {title} {\bibinfo {title} {High-{Intensity}
  {Kapitza}-{Dirac} {Effect}},\ }\href
  {https://doi.org/10.1103/PhysRevLett.61.1182} {\bibfield  {journal} {\bibinfo
   {journal} {Physical Review Letters}\ }\textbf {\bibinfo {volume} {61}},\
  \bibinfo {pages} {1182} (\bibinfo {year} {1988})}\BibitemShut {NoStop}%
\bibitem [{\citenamefont {Ruskin}\ \emph {et~al.}(2013)\citenamefont {Ruskin},
  \citenamefont {Yu},\ and\ \citenamefont {Grigorieff}}]{Ruskin2013}%
  \BibitemOpen
  \bibfield  {author} {\bibinfo {author} {\bibfnamefont {R.~S.}\ \bibnamefont
  {Ruskin}}, \bibinfo {author} {\bibfnamefont {Z.}~\bibnamefont {Yu}},\ and\
  \bibinfo {author} {\bibfnamefont {N.}~\bibnamefont {Grigorieff}},\ }\bibfield
   {title} {\bibinfo {title} {Quantitative characterization of electron
  detectors for transmission electron microscopy},\ }\href
  {https://doi.org/10.1016/j.jsb.2013.10.016} {\bibfield  {journal} {\bibinfo
  {journal} {Journal of structural biology}\ }\textbf {\bibinfo {volume}
  {184}},\ \bibinfo {pages} {385} (\bibinfo {year} {2013})},\ \bibinfo {note}
  {24189638[pmid]}\BibitemShut {NoStop}%
\bibitem [{\citenamefont {Spence}(2013)}]{spence2013high}%
  \BibitemOpen
  \bibfield  {author} {\bibinfo {author} {\bibfnamefont {J.~C.}\ \bibnamefont
  {Spence}},\ }\href@noop {} {\emph {\bibinfo {title} {High-resolution electron
  microscopy}}}\ (\bibinfo  {publisher} {OUP Oxford},\ \bibinfo {year}
  {2013})\BibitemShut {NoStop}%
\bibitem [{\citenamefont {Lupini}(2011)}]{Lupini2011}%
  \BibitemOpen
  \bibfield  {author} {\bibinfo {author} {\bibfnamefont {A.~R.}\ \bibnamefont
  {Lupini}},\ }\bibinfo {title} {The electron ronchigram},\ in\ \href
  {https://doi.org/10.1007/978-1-4419-7200-2_3} {\emph {\bibinfo {booktitle}
  {Scanning Transmission Electron Microscopy: Imaging and Analysis}}},\
  \bibinfo {editor} {edited by\ \bibinfo {editor} {\bibfnamefont {S.~J.}\
  \bibnamefont {Pennycook}}\ and\ \bibinfo {editor} {\bibfnamefont {P.~D.}\
  \bibnamefont {Nellist}}}\ (\bibinfo  {publisher} {Springer New York},\
  \bibinfo {address} {New York, NY},\ \bibinfo {year} {2011})\ pp.\ \bibinfo
  {pages} {117--161}\BibitemShut {NoStop}%
\bibitem [{\citenamefont {Kanters}(2011)}]{Kanters2011}%
  \BibitemOpen
  \bibfield  {author} {\bibinfo {author} {\bibfnamefont {J.~H.~M.}\
  \bibnamefont {Kanters}},\ }\emph {\bibinfo {title} {Electron bunch length
  measurement using the ponderomotive force of a laser standing wave}},\
  \href@noop {} {Master's thesis},\ \bibinfo  {school} {Eindhoven University of
  Technology} (\bibinfo {year} {2011})\BibitemShut {NoStop}%
\bibitem [{\citenamefont {Glaeser}\ \emph {et~al.}(2007)\citenamefont
  {Glaeser}, \citenamefont {Downing}, \citenamefont {DeRosier}, \citenamefont
  {Chiu},\ and\ \citenamefont {Frank}}]{ucb.b1269885920070101}%
  \BibitemOpen
  \bibfield  {author} {\bibinfo {author} {\bibfnamefont {R.~M.}\ \bibnamefont
  {Glaeser}}, \bibinfo {author} {\bibfnamefont {K.}~\bibnamefont {Downing}},
  \bibinfo {author} {\bibfnamefont {D.}~\bibnamefont {DeRosier}}, \bibinfo
  {author} {\bibfnamefont {W.}~\bibnamefont {Chiu}},\ and\ \bibinfo {author}
  {\bibfnamefont {J.}~\bibnamefont {Frank}},\ }\href@noop {} {\emph {\bibinfo
  {title} {Electron crystallography of biological macromolecules.}}}\ (\bibinfo
   {publisher} {Oxford University Press},\ \bibinfo {year} {2007})\BibitemShut
  {NoStop}%
\end{thebibliography}%

\providecommand{\noopsort}[1]{}\providecommand{\singleletter}[1]{#1}%

\end{document}


\title{Supplementary Materials for:\\ Observation of the Relativistic Reversal \\of the Ponderomotive Potential}
\author{Jeremy J. Axelrod, Sara L. Campbell, Osip Schwartz,\\ Carter Turnbaugh, Robert M. Glaeser, Holger M{\"u}ller}
\date{}
\maketitle

\tableofcontents

\newpage

\section{Quasiclassical theory}
To express the phase shift of Eq. \xeqn{eqn:phasecomoving}~ in terms of the laboratory frame Coulomb gauge vector potential, we perform a Lorentz transformation and then restore the Coulomb gauge by a gauge transformation. The Lorentz transformation of the field can be written as
\begin{align}
    0 &= \frac{\gamma}{c} \tilde{\Phi}\left(\mathbf{x},t\right) - \gamma \beta \tilde{A}_z\left(\mathbf{x},t\right) \,, \label{sm:eqn:lt1}\\
    \mathbf{A}'_\perp\left(\mathbf{x}',t'\right) &= \tilde{\mathbf{A}}_\perp\left(\mathbf{x},t\right) \,, \\
    A'_z \left(\mathbf{x}',t'\right) &= - \frac{\gamma}{c} \beta \tilde{\Phi} \left(\mathbf{x},t\right)  + \gamma \tilde{A}_z \left(\mathbf{x},t\right) = \frac{1}{\gamma} \tilde{A}_z \left(\mathbf{x},t\right) \,,
\end{align}
where $\tilde{A}_{z}$ and $\tilde{\mathbf{A}}_{\perp}$ are the longitudinal and transverse components of the laboratory frame vector potential and $\tilde{\Phi}$ is the laboratory frame scalar potential. Fields not in Coulomb gauge are denoted by a tilde. We then restore the Coulomb gauge in the laboratory frame by a gauge transformation:
\begin{align}
    \tilde{\mathbf{A}} &\rightarrow \mathbf{A} = \tilde{\mathbf{A}} + \nabla G  \,, \label{eqn:gaugecondA} \\ 
    \tilde{\Phi} &\rightarrow \Phi = \tilde{\Phi} - \partial_{t} G  = 0 \,, \label{eqn:gaugecond}
\end{align}
where, using Eq. \eqref{sm:eqn:lt1} and Eq. \eqref{eqn:gaugecondA}, we see that Eq. \eqref{eqn:gaugecond} is satisfied when
\begin{align}
    \left[ \partial_t + c\beta \partial_z\right] G\left(\mathbf{x},t\right) &= c \beta A_z \left(\mathbf{x},t\right) \,. \label{sm:eqn:gfun}
\end{align}
This partial differential equation can be directly integrated to give a solution for the gauge function, which is defined for all $\mathbf{x}$ and $t$:
\begin{align}
    G\left(\mathbf{x},t\right) &= c \beta \int_{-\infty}^{t} dT\, A_z \left( \mathbf{x} - c\beta \left(t - T\right) \hat{z} , T \right) \,.
\end{align}
We can then rewrite Eq. \xeqn{eqn:phasecomoving}~ in the laboratory frame in Coulomb gauge:
\begin{align}
    \phi &= - \frac{e^2}{2 m \hbar} \int dt' \, \mathbf{A}'^2\left(\mathbf{r}'_0,t'\right) \,, \\
    &= - \frac{e^2}{2 m \hbar} \int \frac{dt}{\gamma} \, \left[ \tilde{\mathbf{A}}_\perp^2 \left(\mathbf{r}_0\left(t\right),t\right)  + \frac{1}{\gamma^2} \tilde{A}_z^2 \left(\mathbf{r}_0\left(t\right),t\right)  \right] \,, \\
    &=  - \frac{e^2}{2 m \hbar} \int \frac{dt}{\gamma} \, \left[ \tilde{\mathbf{A}}^2 \left(\mathbf{r}_0\left(t\right),t\right)  - \beta^2 \tilde{A}_z^2 \left(\mathbf{r}_0\left(t\right),t\right)  \right] \,, \\
    &= - \frac{e^2}{2m\hbar}  \int \frac{dt}{\gamma} \, \left[  \left( \mathbf{A} \left(\mathbf{r}_0\left(t\right),t\right) - \nabla G \left(\mathbf{r}_0\left(t\right),t\right)  \right)^2 \right.\\
    &- \beta^2 \left. \left( A_z \left(\mathbf{r}_0\left(t\right),t\right) - \nabla_z G \left(\mathbf{r}_0\left(t\right),t\right)  \right)^2    \right]  \,.
\end{align}

Taking the vector potential from Eq. \xeqn{eqn:vectorpotential}~ and assuming that the envelope $A_0\left(y,z\right)$ varies slowly relative to the wave cycle along the electron trajectory, we may approximate
\begin{align}
    G\left(\mathbf{x},t\right) &\cong \frac{c \beta}{\omega} \cos\left(\theta\right) A_0 \left(y,z\right) \cos\left( 2 \pi x / \lambda_L\right)  \sin\left(\omega t\right) \,,
\end{align}
where 
\begin{align}
    \nabla G\left(\mathbf{x},t\right) &\cong  - \hat{x} \beta \cos\left(\theta\right) A_0 \left(y,z\right) \sin\left(2 \pi x / \lambda_L\right) \sin\left(\omega t\right) \,,
\end{align}
so that when averaging over the rapidly oscillating components, we find that
\begin{align}
    \phi\left(x,y\right) &= - \frac{1}{\hbar} \int dt \,  \frac{e^2 A_0^2\left(y,z(t)\right)}{4 m \gamma} \frac{1}{2} \left[ 1 + \left(1- 2 \beta^2 \cos^2\left(\theta\right) \right) \cos\left(4 \pi x / \lambda_L\right)   \right] \,, \\
    &= - \frac{1}{\hbar} \int dz \,  \frac{e^2 A_0^2\left(y,z\right)}{4 m c \beta \gamma} \frac{1}{2} \left[ 1 + \left(1- 2 \beta^2 \cos^2\left(\theta\right) \right) \cos\left(4 \pi x / \lambda_L\right)   \right] \,.
\end{align}

\section{Polarimeter calibration}
\subsection{Polarimeter Design}

\begin{figure}[H]
\centering
\includegraphics[width = 5in]{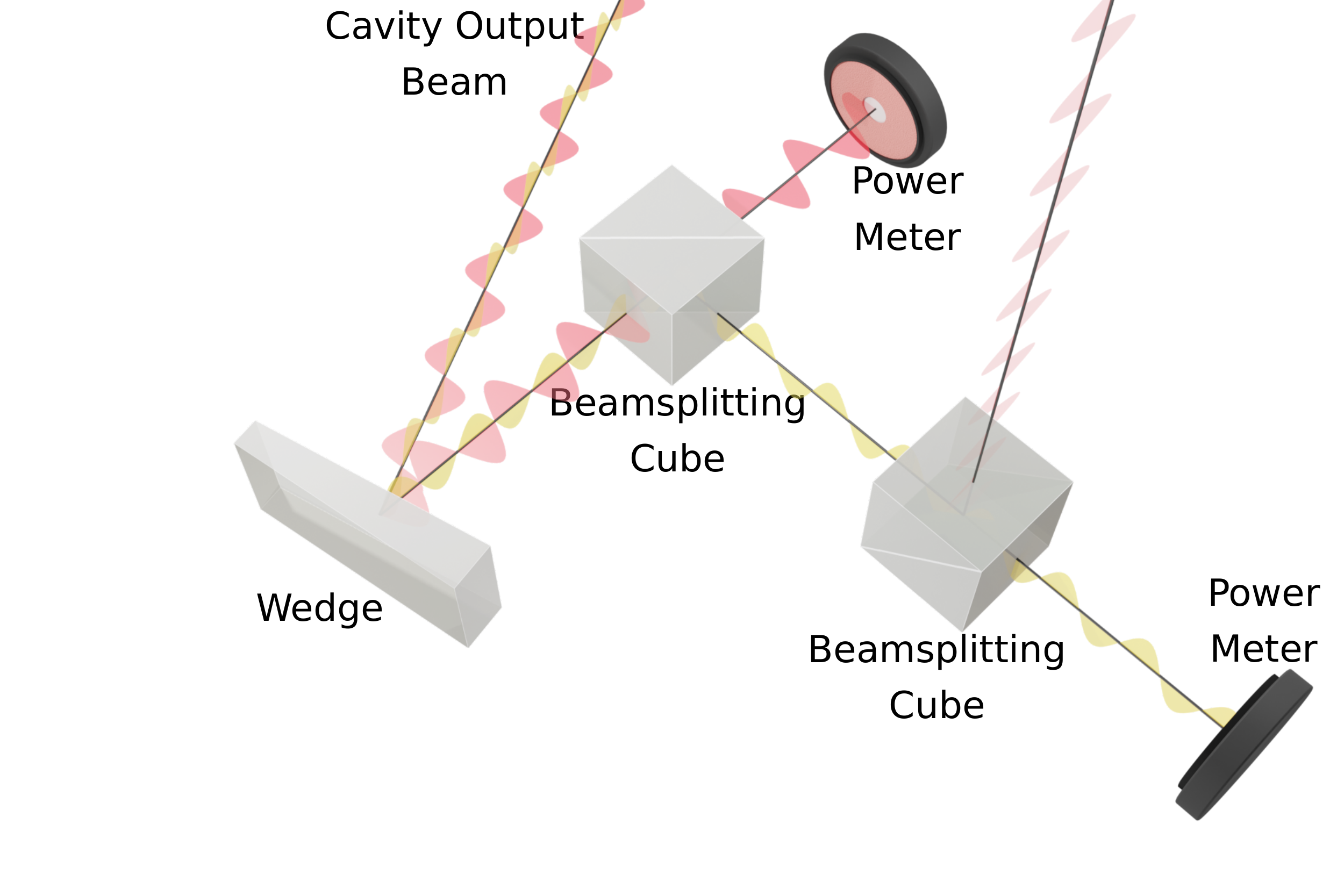}
\caption{Diagram of the polarimeter.
Yellow is used to mark vertically polarized light (relative to the polarimeter axis)
and red is used to mark horizontally polarized light.}
\label{fig:polarimeter}
\end{figure}

The polarimeter is installed at the output of the cavity, as shown in Fig. \xfig{schematic}~. In order to reduce the optical power to within the range acceptable to the power meters used ($0-500\,\mathrm{mW}$),
the first optic in the polarimeter is a 10\% reflectance, $5^\circ$ beam sampling wedge
which the beam hits near normal incidence.
There are then two polarizing beamsplitting cubes and two $1064\,\mathrm{nm}$ optical power meters.
The first cube is oriented such that both the transmitted and reflected beams exit parallel to the optical breadboard.
This establishes the axis of the polarimeter to be nearly parallel to the optical axis of the electron microscope, although there could remain a mechanical misalignment
between the axes or linear rotation of the polarization by cavity output optics
causing a disagreement in the polarization angle at the electron beam and the polarimeter (denoted $\xi$ in the main text).
The transmitted beam from the first cube is then directed into one of the optical power meters.
The reflected beam passes through a second polarizing beamsplitting cube,
which is oriented such that the polarization transmitted through the second
is the same polarization that is reflected by the first cube.
The second cube serves to increase the extinction ratio of the polarimeter;
the polarizing beamsplitting cubes have an extinction ratio better than $1000 : 1$
for p-polarized : s-polarized light in the transmitted beam,
but only approximately $100 : 1$ for s-polarized : p-polarized light in the reflected beam.
The second optical power meter is placed after the second cube.
The setup is shown in Fig. \ref{fig:polarimeter}.

\subsection{Theory of Calibration}

Here we develop a model of the polarimeter's response to linearly polarized light,
which was used to determine the polarization angle during experiments.
Let the electric field incident to the polarimeter be $\bm{E}_0$.
With $\eta$ being the angle from vertical 
and $\psi$ the relative phase between the vertical and horizontal components of the light,
the input electric field can be written generally as a Jones vector:
\begin{equation}
    \bm{E}_0
    = \begin{pmatrix}
        \sin(\eta) \\ \cos(\eta) e^{i \psi}
    \end{pmatrix} \,.
    \label{equ:E0def}
\end{equation}
We ignore the phase of the horizontal component,
as only the relative phase of the two components is measurable.

The polarimeter measures the intensity of light at the two power meters.
Let the electric fields at the power meters be $\bm{E}_1$ and $\bm{E}_2$,
which can be written in terms of the input electric field
and a Jones matrix $S_{1,2}$ describing the transfer function from the input of the polarimeter to each power meter as
\begin{equation}
    \bm{E}_1 = S_1 \bm{E}_0 \text{ and }
    \bm{E}_2 = S_2 \bm{E}_0 \,.
\end{equation}

In an ideal polarimeter, only light of one polarization should be transmitted to each power meter.
However, due to the finite extinction ratio of the polarizing beamsplitting cubes,
we expect a small fraction $\delta_{1,2}$ of the light of the other polarization to reach each power meter.
Some light is also lost in the polarimeter.
Reflection from the wedge is polarization dependent,
so this loss factor $T_{1,2}$ could be different for the two polarizations.
Therefore, the Jones matrices should be
\begin{equation}
    S_1 = T_1 \begin{pmatrix} \delta_1 & 0 \\ 0 & 1 \end{pmatrix} \text{ and }
    S_2 = T_2 \begin{pmatrix} 1 & 0 \\ 0 & \delta_2 \end{pmatrix}\,.
    \label{equ:Tjdef}
\end{equation}
We assume the polarization is uniform throughout the beam
and the beam intensity profile does not depend on the polarization or phase.
This introduces an additional proportionality factor, which is incorporated into $T_1$ and $T_2$.
Thus, the power at each power meter is
\begin{equation}
    P_{1,2} = \left| S_{1,2} \bm{E}_0 \right|^2 \,.
\end{equation}

Using Eq. \eqref{equ:E0def} and Eq. \eqref{equ:Tjdef}, we expand the expressions for the power to get
\begin{equation}
    P_1
    = \left| S_1 \bm{E}_0 \right|^2
    = \left| E_0 T_1 \right|^2
    \left| \begin{pmatrix} \delta_1 \sin^2(\eta)\\ \cos^2(\eta) \end{pmatrix} \right|^2
    = \left| E_0 T_1 \right|^2 \left( \delta_1^2 \sin^2(\eta) + \cos^2(\eta) \right)
\end{equation}
and
\begin{equation}
    P_2
    = \left| S_2 \bm{E}_0 \right|^2
    = \left| E_0 T_2 \right|^2
    \left| \begin{pmatrix} \sin(\eta)\\ \delta_2 \cos(\eta) \end{pmatrix} \right|^2
    = \left|E_0 T_2 \right|^2 \left( \sin^2(\eta) + \delta_2^2 \cos^2(\eta) \right)\,.
\end{equation}
We use the ratio of the detected powers $R$ to determine the polarization independent of the input power:
\begin{equation}
    R = \frac{P_1}{P_2}
    = X_1 \cdot \frac{\cos^2(\eta) + X_2 \sin^2(\eta)}{\sin^2(\eta) + X_3 \cos^2(\eta)}\,,
    \label{sm:equ:R_model}
\end{equation}
where
\begin{equation}
    X_1 = \frac{\left| T_1 \right|^2}{\left| T_2 \right|^2} \,,
    ~~~
    X_2 = \delta_1^2 \,,
    ~~~\text{and}~~~
    X_3 = \delta_2^2 \,.
\end{equation}

\subsection{Calibration Procedure}

\begin{figure}[H]
\centering
\includegraphics[width = 5in]{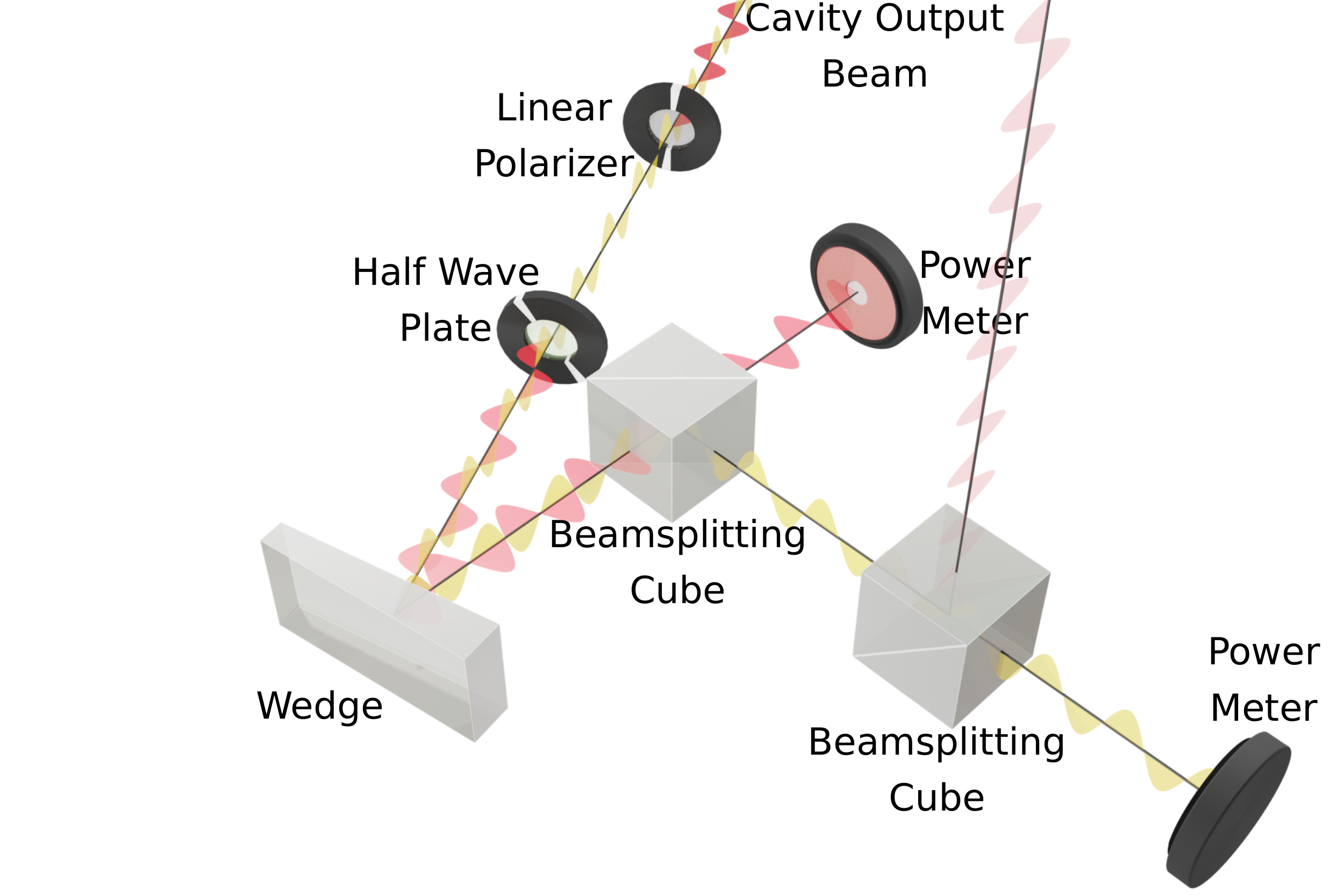} 
\caption{Diagram of the optical setup during polarimeter calibration.
In addition to the previous polarimeter optics,
a linear polarizer is added at the cavity output to produce linearly polarized light.
The transmission angle of the linear polarizer was not referenced 
to either the polarimeter or electron microscope axes.
After the linear polarizer, a half-wave plate was installed.
During calibration the half-wave plate was rotated in order to produce the range of linear polarizations.}
\label{fig:polarimeter_calib}
\end{figure}

In order to determine $R$ as a function of $\eta$, the optical power at each of the power meters was measured for a range of linear polarization angles. The range of linear polarization angles
was produced by placing a linear polarizer followed by a half-wave plate
between the output of the cavity and the polarimeter, and then rotating the half-wave plate. The calibration relies on the beam intensity profile remaining constant regardless of angle,
so it was necessary to use the same alignment for calibration that would be used during data collection.
It was found that rotating a linear polarizer shifted the position of the beam,
so the half-wave plate was rotated while the linear polarizer remained fixed.
The calibration setup is shown in Fig. \ref{fig:polarimeter_calib}.

After simultaneously measuring $R$ and $\eta$,
the $X$ parameters were determined by a nonlinear least squares fit of the data to Eq. \ref{sm:equ:R_model}.
The polarization axis of the linear polarizer was not aligned relative to the polarimeter,
so the value of $\eta$ in the data was not determined absolutely.
Therefore, an angle offset $\eta_0$ was included in the fit.
To perform the fit, the logarithm of Eq. \ref{sm:equ:R_model} was fit to the logarithm of the measured
$R$ versus $\eta$, as the measured values of $R$ spanned several orders of magnitude.

\subsection{Calibration Results}

\begin{figure}[H]
\centering
\includegraphics[width = 5in]{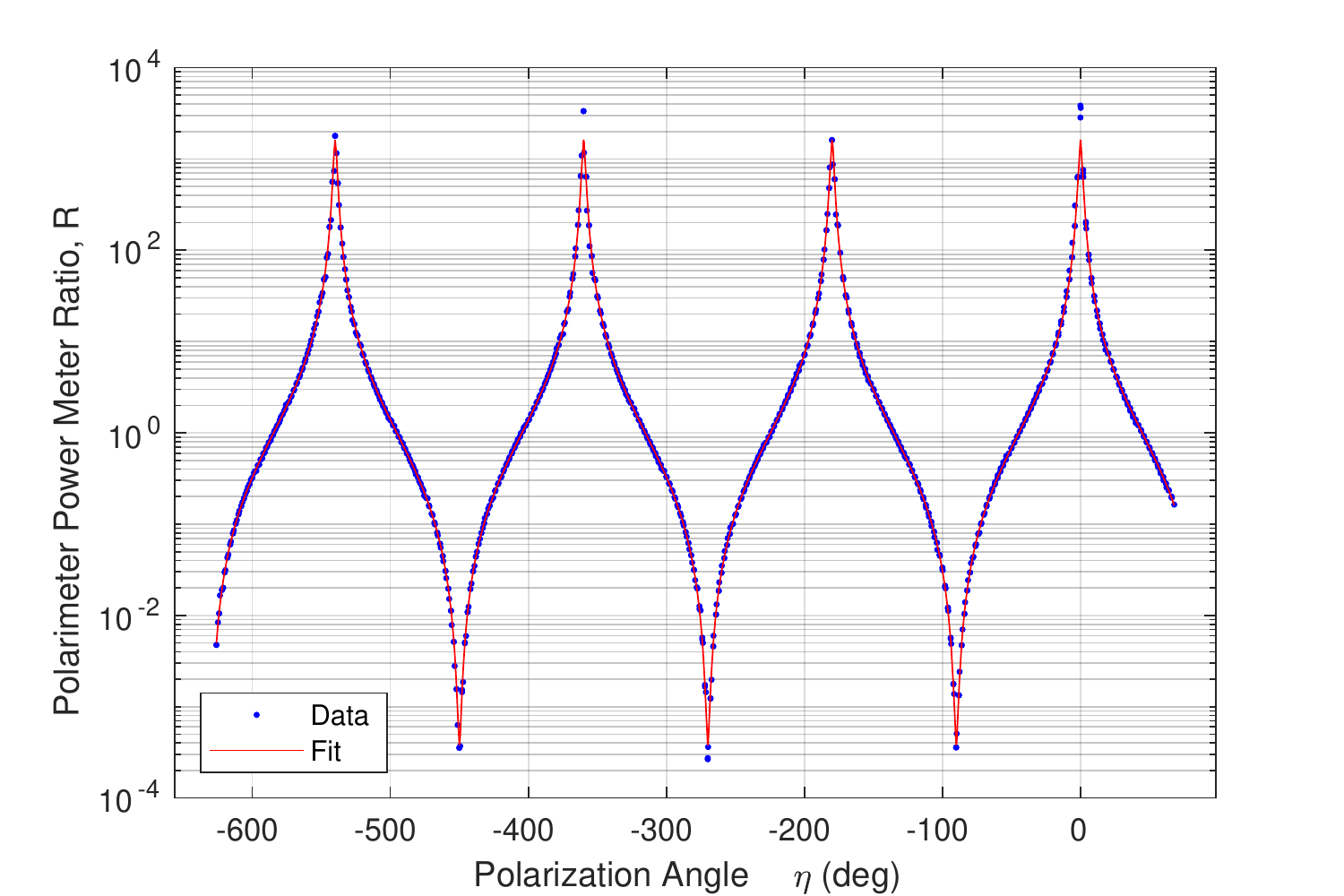} 
\caption{Polarimeter calibration results. The blue dots are measured values of $R$
and the red line is the fit.
The fit is in good agreement with the data,
suggesting that the parameters are an accurate model 
for the polarimeter's response to light of various linear polarizations.}
\label{fig:calib_results}
\end{figure}

\begin{figure}[H]
\centering
\includegraphics[width=5in]{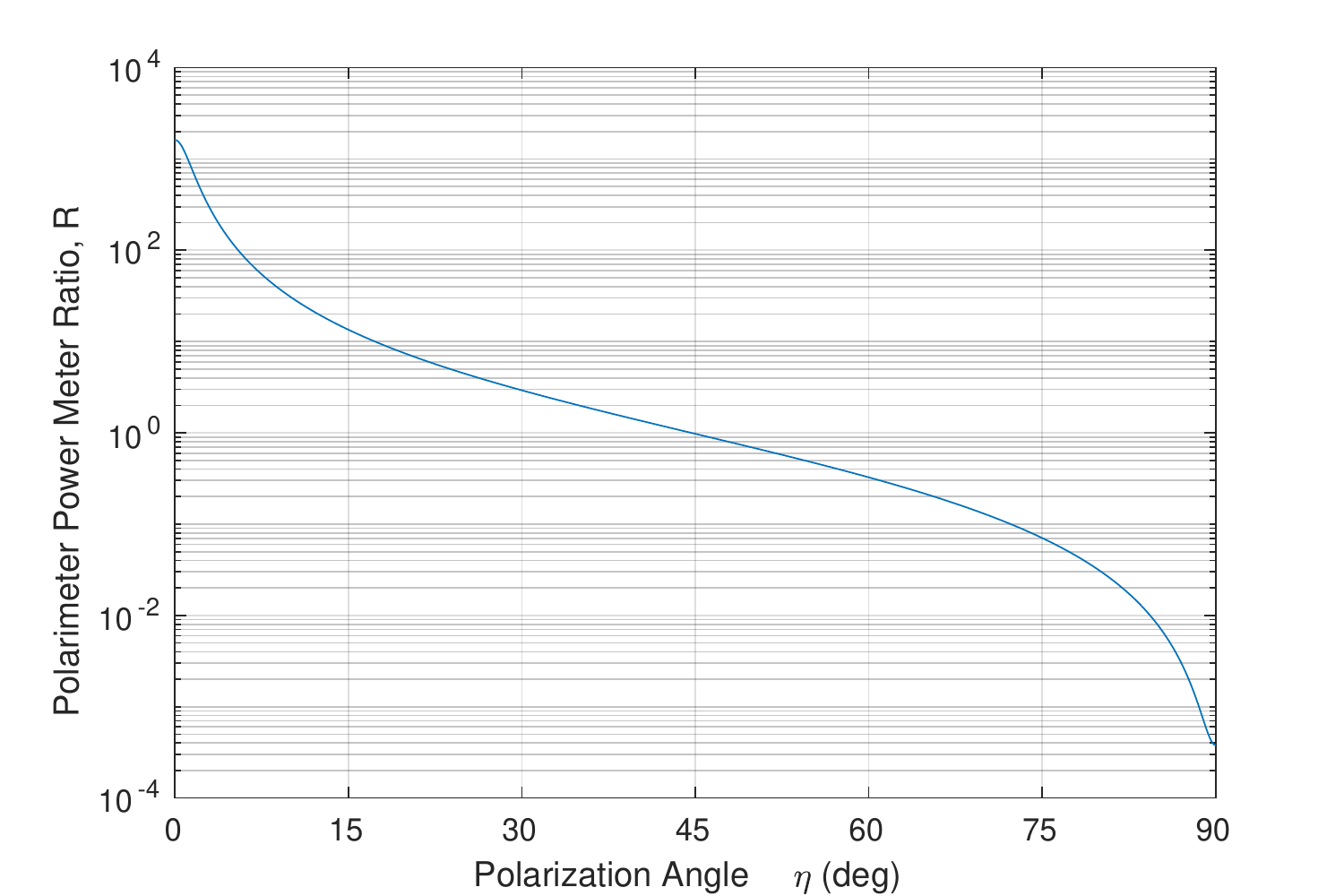} 
\caption{A single $0^\circ$--$90^\circ$ range of the fit ratio as a function of angle. This function was used to convert measured polarimeter power meter ratios to polarization angles.}
\label{fig:calib_results_single}
\end{figure}

The extinction ratios for the polarimeter were both found to be greater than $1000 : 1$
for light that was both horizontally and vertically polarized relative to the polarimeter,
as shown in Fig. \ref{fig:calib_results}.
Parameters were found which fit the function to the measured calibration curve:
\begin{equation}
    X_1 = 9.736 \times 10^{-1} \,,~
    X_2 = 3.892 \times 10^{-4} \,,~
    X_3 = 6.0461 \times 10^{-4} \,.
\end{equation}
Note that the parameters besides $X_1$ are small,
suggesting that the polarimeter behaves similarly to an ideal polarimeter.
After fitting, the fit function was inverted for a single interval of $0^\circ \leq \eta \leq 90^\circ$
in order to determine the the polarization angle from the ratio.
The fit function over this interval is shown in Fig. \ref{fig:calib_results_single}.
During the experiment, all measured ratios were within the domain of the fit function.

\subsection{Polarimeter accuracy}
\begin{figure}[H]
\centering
\includegraphics[width = 5in]{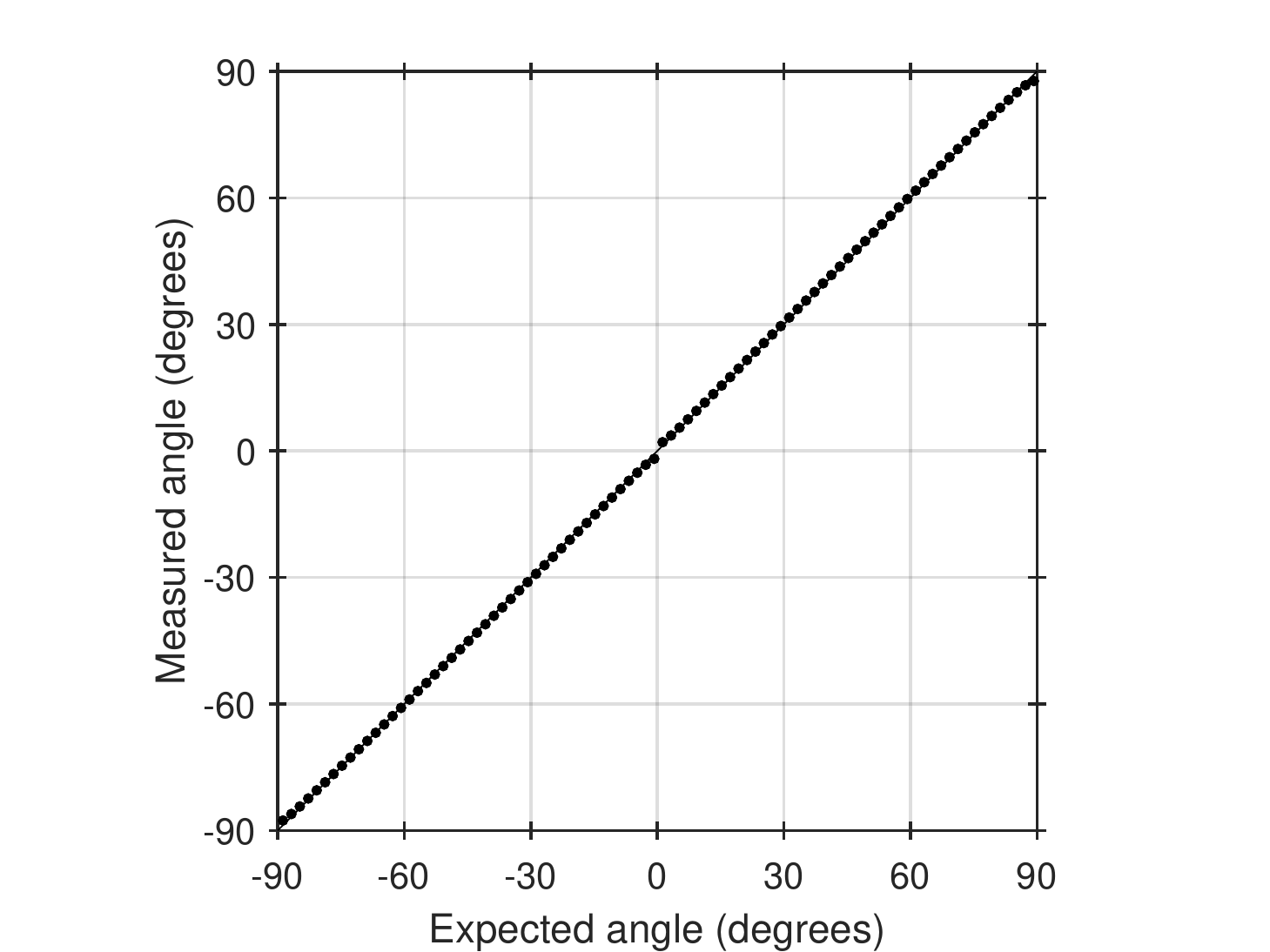}
\caption{The polarization angle measured by the polarimeter as a function of the polarization angle expected from the input half-wave plate position.}
\label{sm:fig:polacc1}
\end{figure}

\begin{figure}[H]
\centering
\includegraphics[width = 5in]{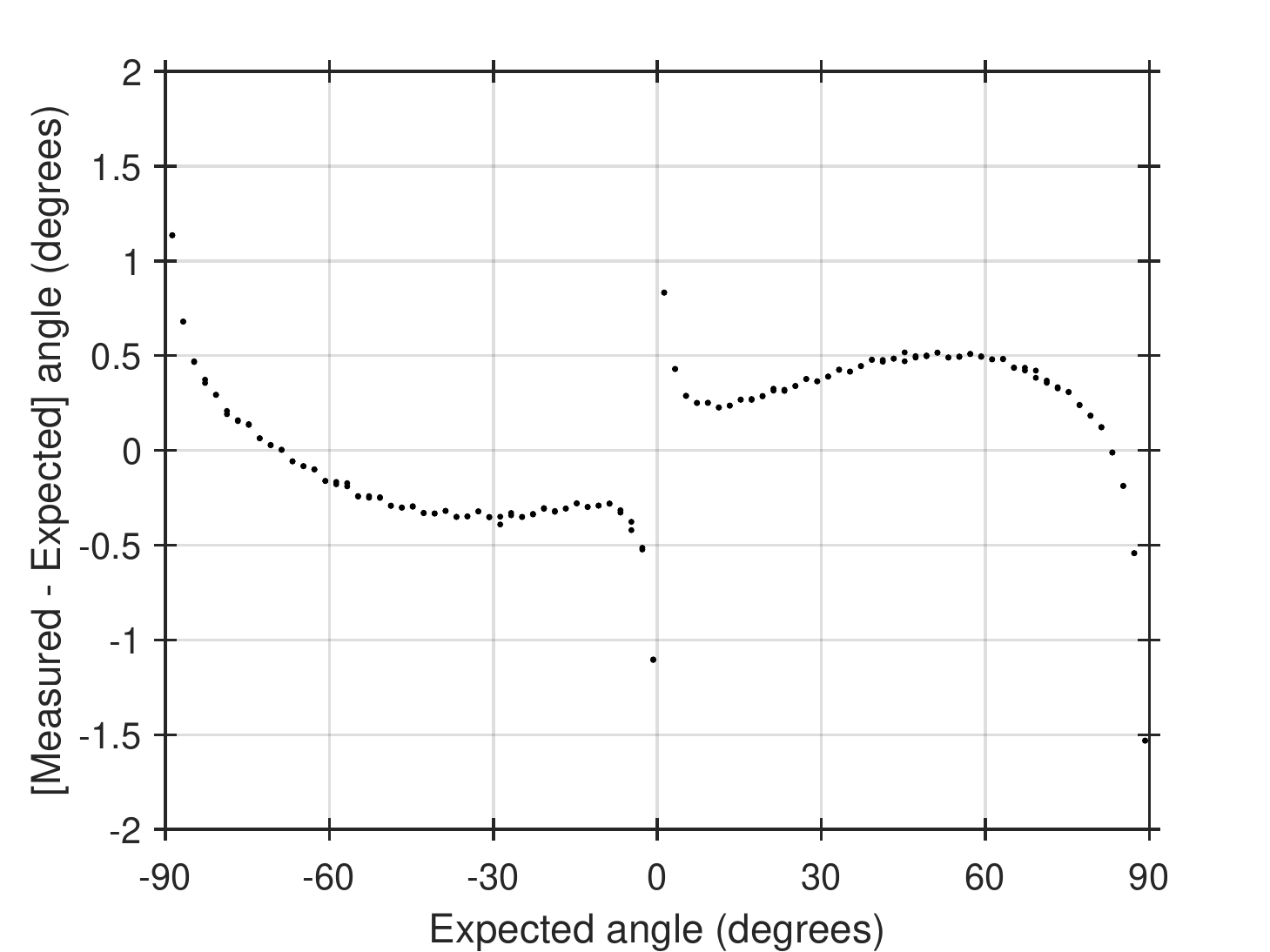}
\caption{The difference between the measured angle and the expected angle is shown as a function of the expected angle.}
\label{sm:fig:polacc2}
\end{figure}

The polarimeter's measurement will only accurately represent the polarization of light inside the optical cavity if the polarization is not substantially changed between the inside of the cavity and the input of the polarimeter. In our experiment, there were three optical elements between the cavity and polarimeter input: the cavity output mirror, a beam deflection prism with a $18.15^\circ$ wedge angle, and a vacuum window. All surfaces were anti-reflection coated. We were unable to directly measure the intra-cavity polarization, so instead we compare the polarization angle measured by the polarimeter with the polarization angle expected given the angle of the half-wave plate at the input of the cavity. The change in input polarization angle is expected to be twice that of the change in half-wave plate angle; the zero of the expected angle is defined to be where a linear fit through the measured angle data crosses zero. This comparison, shown in Fig. \ref{sm:fig:polacc1}, characterizes the polarization distortion induced by all optics between the input half-wave plate and polarimeter. In addition to the optical elements listed above, this includes the input vacuum window, the input cavity mirror, and the possible distortion induced by the small frequency difference between the cavity polarization eigenmodes. It therefore likely underestimates the polarimeter's accuracy. The difference between the measured and expected angles is shown in Fig. \ref{sm:fig:polacc2}. The difference is less than 2 degrees for all positions of the input half-wave plate, which is a rough upper limit on the polarimeter's accuracy.

\section{Data analysis}
\subsection{Ronchigram fitting}
In this section we develop a model relating the electron beam phase modulation profile to the Ronchigram formed in the image plane. This model was used to determine the phase modulation depth for each Ronchigram. 

As derived in the supplementary materials of \xcite{schwartz_laser_2018}~, the electron wavefunction in the image plane due to a spatial phase modulation $\phi\left(\bm{x}\right)$ applied a distance $\Delta$ beyond the diffraction (focal) plane is
\begin{align}
\psi_{im}\left(\bm{x}\right) &\propto \int d^2\bm{x}_1 \, h \left( \frac{\Delta}{Mf} \bm{x} - \bm{x}_1 ; \Delta\right) \cdot e^{-i \phi\left(\bm{x}_1\right)} \,,  \label{sm:eqn:psiim}\\
h\left(\bm{x};z\right) &\equiv \frac{1}{2\pi} \frac{ k e^{ikz}}{i z } e^{i\frac{k}{2z} \left|\bm{x}\right|^2} \,, ~~~~\bm{x} \equiv \left(x,y\right) \,,
\end{align}
where $M$ is the microscope magnification, $f$ is the effective focal length of the microscope's objective lens, and $k$ is the electron's angular wavenumber. Using Eq. \xeqn{eqn:pprel}~ and the fact that the numerical aperture $\mathrm{NA} = \frac{\lambda_L}{\pi w_0}$ of the cavity mode is small, we can write that
\begin{align}
    \phi\left(x,y\right) &\cong \bar{\phi} e^{-\frac{1}{2} \left(\mathrm{NA} \cdot k_L y\right)^2} \cdot \frac{1}{2} \left[1 + e^{-2 \left(\frac{\Theta}{\mathrm{NA}}\right)^2 } \rho\left(\theta,\beta\right) \cos\left( 2 k_L x\right) \right] \,,
\end{align}
where $\bar{\phi}$ is the mean phase shift along the $x$-direction at $y=0$, $w_0$ is the cavity mode waist, $\Theta$ is the angle between the cavity mode axis and the electron beam axis ($\hat{z}$) minus $\pi/2 \, \mathrm{rad}$, and $k_L = 2\pi/\lambda_L$ is the laser angular wavenumber. 

Inserting this expression into Eq. \eqref{sm:eqn:psiim}, using the Jacobi-Anger identity, and integrating under the assumption that $\mathrm{NA} \ll 1$ gives
\begin{align}
    \psi_{im}\left(x,y\right) &\propto \sum_{n=-\infty}^{\infty}   e^{i \left(-\pi n /2 - 2 \frac{\Delta k_L^2}{k}  n^2\right)}        e^{i 2 n  \frac{\Delta}{M f} k_L x}   \cdot        J_n\left(  \frac{\bar{\phi}}{2} e^{- \frac{1}{2}\left( \mathrm{NA} \cdot    \frac{\Delta}{M f} k_L y\right)^2} e^{- 2 \left(\frac{\Theta}{\mathrm{NA}}\right)^2 } \rho\left(\theta,\beta\right)  \right) \,.
\end{align}

The resulting intensity distribution is therefore
\begin{align}
    I\left(x,y\right) &= \left| \psi_{im}\left(x,y\right)\right|^2 \\
    &= \sum_{n=-\infty}^{\infty}   e^{i 2 n \frac{\Delta}{Mf} k_L x} 
\cdot \sum_{m=-\infty}^{\infty} e^{-i \left(\pi n /2  \right)}  e^{-i 2   \frac{\Delta k_L^2}{k} \left(n^2 + 2 m n\right) } \nonumber \\
&\cdot J_{n+m}\left( \frac{\bar{\phi}}{2} e^{- \frac{1}{2}\left( \mathrm{NA} \cdot    \frac{\Delta}{M f} k_L y\right)^2} e^{- 2 \left(\frac{\Theta}{\mathrm{NA}}\right)^2 } \rho\left(\theta,\beta\right)   \right) \cdot J_{m}\left( \frac{\bar{\phi}}{2} e^{- \frac{1}{2}\left( \mathrm{NA} \cdot    \frac{\Delta}{M f} k_L y\right)^2} e^{- 2 \left(\frac{\Theta}{\mathrm{NA}}\right)^2 } \rho\left(\theta,\beta\right)   \right) \,.
\end{align}
This expression is normalized such that $I=1$ when $\bar{\phi}=0$. The image is expressed in terms of a Fourier series along the $x$-axis, so that the power spectral density in the $n$-th order term (as determined by an $N$-point discrete Fourier transform) is 
\begin{align}
    \left| \tilde{I}_n \left( y \right) \right|^2 &= N^2 \left| \sum_{m=-\infty}^{\infty}  e^{-i 4   \frac{\Delta k_L^2}{k}  n m } \cdot J_{n+m}\left( \frac{\bar{\phi}}{2} e^{- \frac{1}{2}\left( \mathrm{NA} \cdot    \frac{\Delta}{M f} k_L y\right)^2} e^{- 2 \left(\frac{\Theta}{\mathrm{NA}}\right)^2 } \rho\left(\theta,\beta\right)   \right) \right.  \nonumber \\
    &~~~~~~~~~~~~~~~ \left. \cdot J_{m}\left( \frac{\bar{\phi}}{2} e^{- \frac{1}{2}\left( \mathrm{NA} \cdot    \frac{\Delta}{M f} k_L y\right)^2} e^{- 2 \left(\frac{\Theta}{\mathrm{NA}}\right)^2 } \rho\left(\theta,\beta\right)   \right) \right|^2 \,. \label{sm:eqn:psd}
\end{align}

The following procedure was used to fit each Ronchigram:
\begin{enumerate}
    \item Remove dead pixels from the image and normalize the background level to unity.
    \item Determine the value of $\frac{\Delta}{Mf} k_L$ and direction of the $x$-axis. This was done taking a 2D discrete Fourier transform of the image and locating the non-zero spatial frequency at which its magnitude squared was largest.
    \item Interpolate the image onto a grid aligned with the axis of the standing wave in the image.
    \item Perform a 1D discrete Fourier transform of the interpolated image along the standing wave axis and extract the power spectral density of the $n = 1,2$ orders as a function of the transverse coordinate, $y$. 
    \item Fit these power spectral density profiles to the model described by Eq. \eqref{sm:eqn:psd} as a function of the variables $\frac{\Delta k_L^2}{k}$, $\bar{\phi}  \rho\left(\theta,\beta\right)$, and $\mathrm{NA}$.
\end{enumerate}
 For the purposes of fitting, it was assumed that $\Theta=0$, since during the experiment $\Theta$ was held constant and so the factor $e^{-2\left(\frac{\Theta}{\mathrm{NA}}\right)^2}$ was effectively absorbed into the fit's estimation of $\bar{\phi}$, which was then normalized to extract a measurement of $\rho\left(\theta,\beta\right)$. By comparing the measured value of $\bar{\phi}$ to that predicted by Eq. \xeqn{eqn:phi0}~, the tilt angle $\Theta$ was estimated to be between $1.2^\circ$ and $1.6^\circ$. Contamination or defects on the cavity mirror surfaces prevented the mode from being aligned such that $\Theta =0^\circ$ while maintaining sufficient mirror reflectivity to support the required circulating cavity power.

\subsection{Figure 2(c) fringe position drift} 
A temporally linear drift in the Ronchigrams' fringe position was removed from the data displayed in Fig. \xfig{fringereversal}~. During the course of each experiment, rotating the cavity input half-wave plate caused the cavity input coupling efficiency to change. The concomitant change in circulating cavity power (of approximately 3\%) caused the amount of heat deposited in the cavity optomechanics to change proportionally, leading to thermal expansion of the cavity support structure and overall motion of the cavity and cavity mode (along the laser beam axis) relative to the electron beam.

\begin{figure}[H]
\centering
\includegraphics[width = 5in]{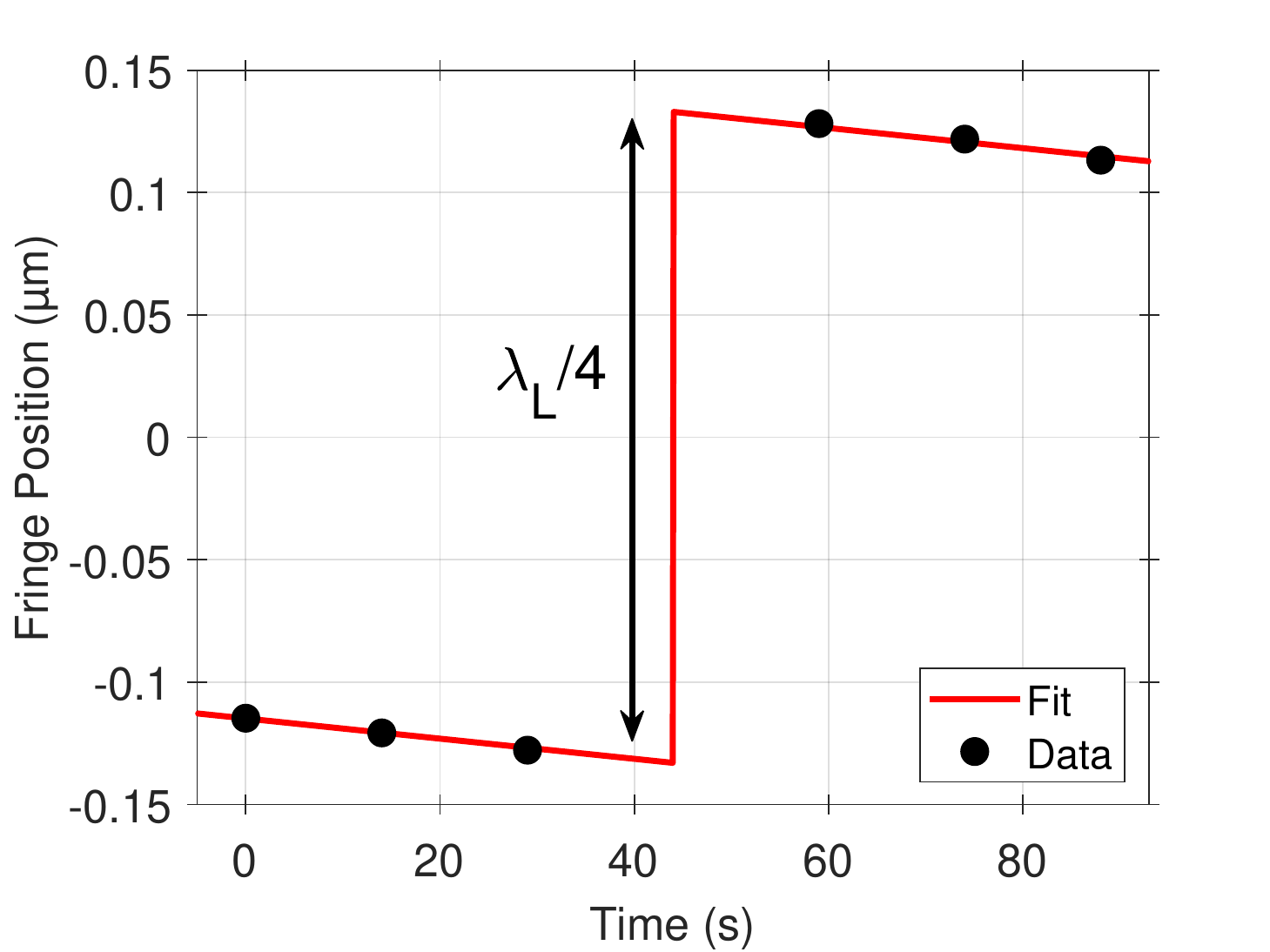} 
\caption{Ronchigram fringe position versus time of Ronchigram capture for the data shown in Fig. 2(c) (black dots). The linear fit (including $\lambda_L/4$ jump across the relativistic reversal angle) is shown in red. The fitted drift rate is $0.4\,\mathrm{nm}/\mathrm{s}$.}
\label{sm:fig:driftcorr}
\end{figure}

The Ronchigram fringe position was plotted as a function of time for the data sets displayed in Fig. \xfig{fringereversal}~ (excluding the one near the relativistic reversal angle), and a line was fit to the data, assuming a shift in fringe position of $\lambda_L/4$ across the relativistic reversal angle. This data is shown in Fig. \ref{sm:fig:driftcorr}. The linear drift was then subtracted from the fringe position data before it was plotted in Fig. \xfig{fringereversal}~.

\subsection{Mode waist measurement}
The waist of the cavity mode $w_0$ was measured by weakly phase modulating the cavity input light in order to generate frequency sidebands. The sideband frequency was then scanned, while the cavity output light was monitored on a photodiode. When one of the sidebands becomes resonant with the $\mathrm{TEM}_{01}$ or $\mathrm{TEM}_{10}$ mode of the cavity, the cavity output light intensity oscillates at the cavity's transverse mode spacing frequency. However, since the transverse mode profiles of the cavity are orthogonal functions, the total power in the beam remains constant. Therefore, to generate a non-zero signal, a portion of the beam was blocked before reaching the photodiode. 

In a Fabry-P{\'e}rot cavity, the transverse mode spacing frequency (non-angular) $\nu$ is related to the mode waist $w_0$ by the formulas
\begin{align}
    w_0 &= \sqrt{ \frac{\lambda_L R}{2 \pi} \sqrt{1 -g^2}} \,, \\
    \nu &= \frac{c}{2 R \left(1-g\right)} \left(1 - \arccos\left(g\right)/\pi\right) \,,
\end{align}
which can be inverted numerically. $\lambda_L$ is the laser wavelength, $R$ is the cavity mirror radius of curvature (symmetric cavity assumed), $g$ is the cavity stability parameter, and $c$ is the speed of light.

This method was used to measure the cavity mode waist instead of relying on fitted values from the Ronchigram fitting because those values, while accurate for larger phase shift values, become inaccurate for small phase shift values where the data being fitted is dominated by shot noise.

\section{Additional data}

\subsection{Polarization dependence}
\begin{figure}[H]
\centering
\includegraphics[width = 5in]{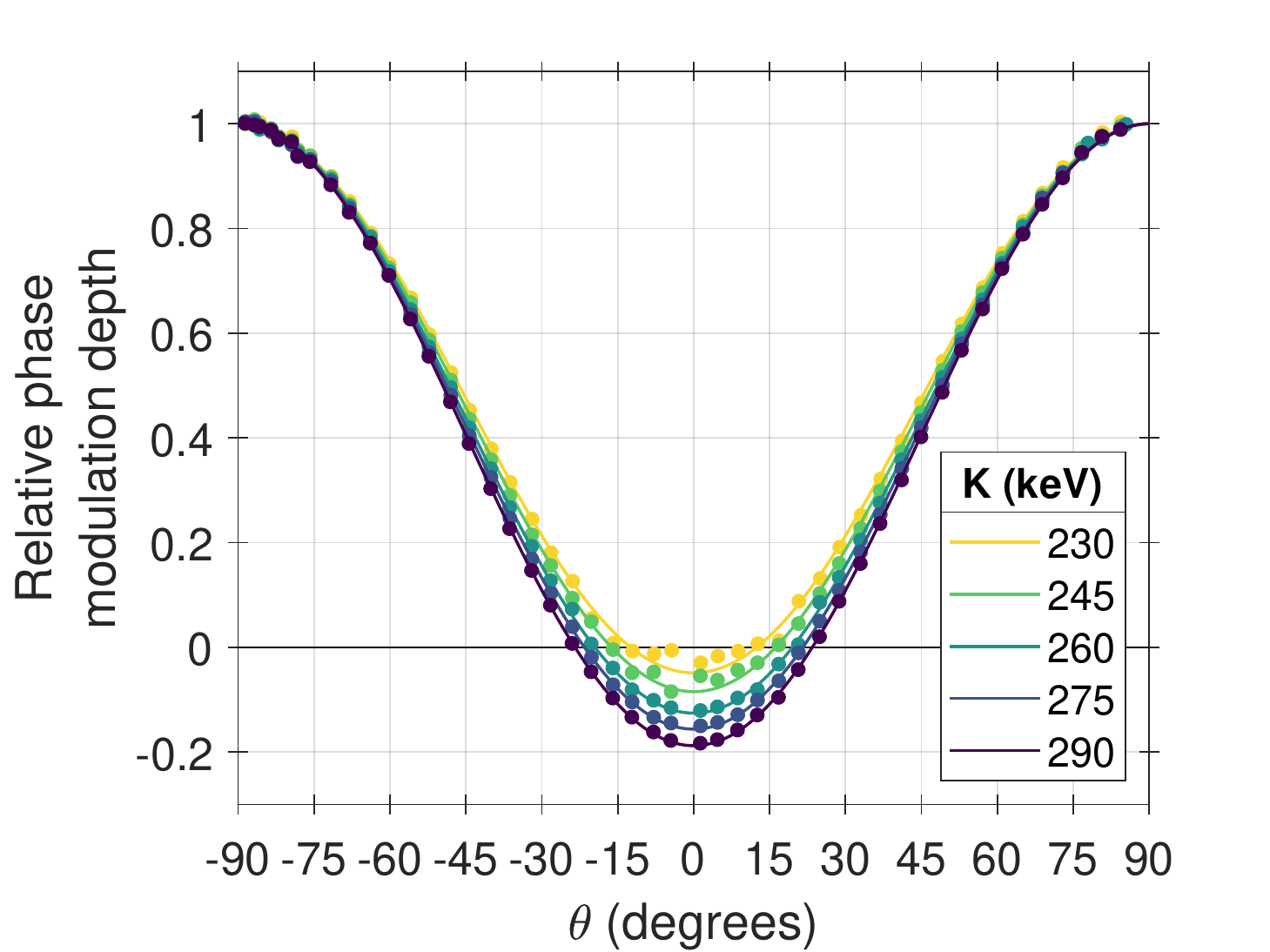} 
\caption{Measurements of the relative phase modulation depth (dots) are plotted along with fitted theory curves (lines) as a function of the laser wave polarization angle $\theta$ for the values of electron energy $K$ (equivalently, electron speed $\beta$) that were omitted from Fig. 2(a) for clarity. The fitted theory (solid lines) is described by Eq. (7).}
\label{sm:fig:others}
\end{figure}

The data sets omitted from Fig. \xfig{angledependence}~ for clarity are shown in Fig. \ref{sm:fig:others}.

\subsection{Electron energy measurement}
\begin{figure}[H]
\centering
\includegraphics[width = 5in]{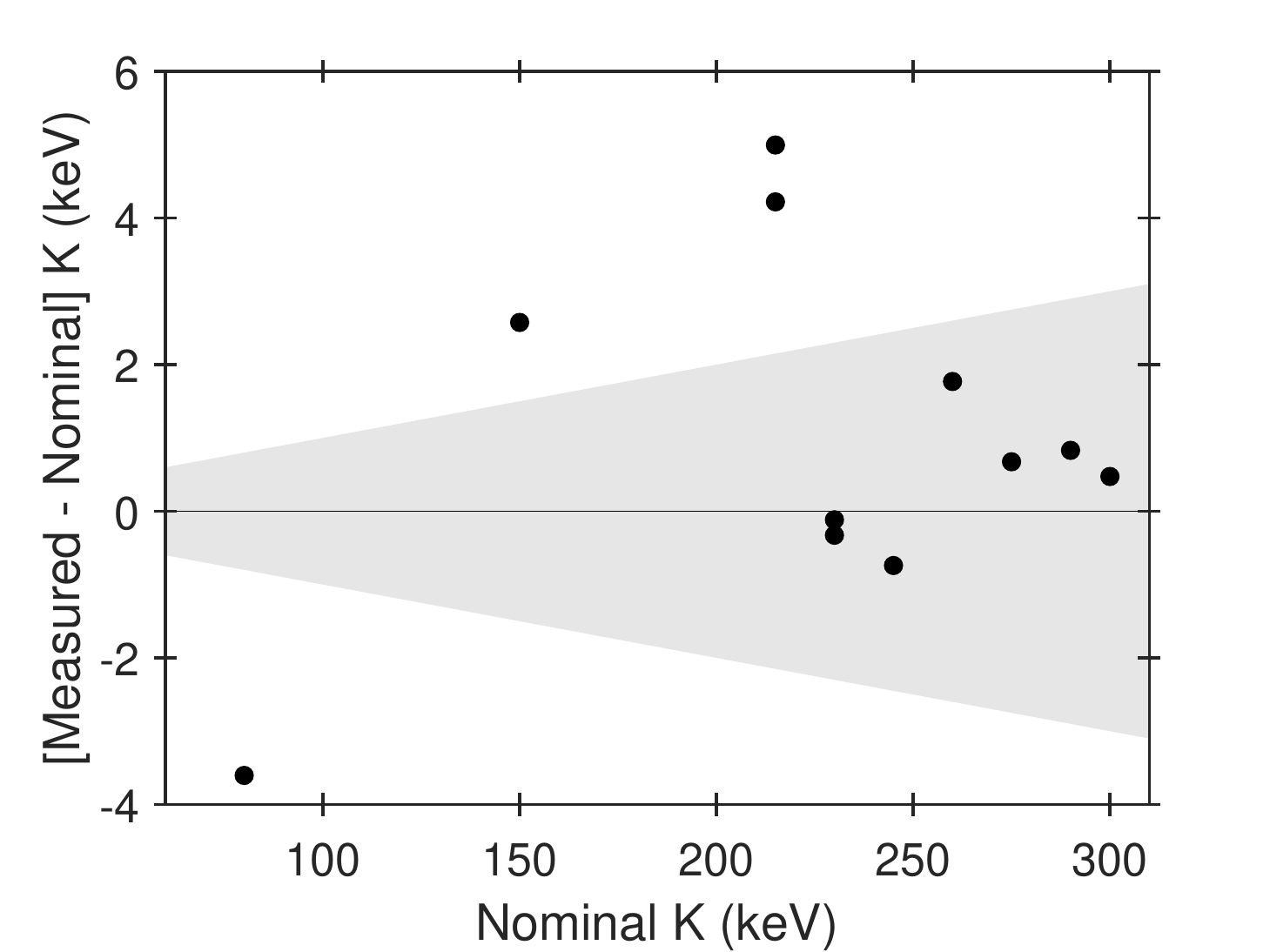} 
\caption{The difference between the fitted and nominal electron energies is shown as a function of the nominal electron energy. The region of $\pm 1 \%$ nominal energy uncertainty is represented by the shaded area.}
\label{sm:fig:ecompare}
\end{figure}

The fit values of the electron speed $\beta$ for the polarization angle dependence curves shown in Fig. \xfig{angledependence}~ and Fig. \ref{sm:fig:others} can be written in terms of the electron energy and compared to the nominal values set by the electron microscope controls. The difference between the measured and nominal energy is shown as a function of nominal energy in Fig. \ref{sm:fig:ecompare}. The region of nominal energy uncertainty as specified by the manufacturer ($\pm 1 \%$) is also shown. The agreement is within the nominal energy uncertainty for nominal energies $\geq 230\, \mathrm{keV}$. For the nominal energy $\leq 215\, \mathrm{keV}$, the difference between the fitted and nominal energy exceeds $\pm 1 \%$ but remains within $\pm 5 \%$ for all energies.

\subsection{Elliptical polarization dependence}
\begin{figure}[H]
\centering
\includegraphics[width = 5in]{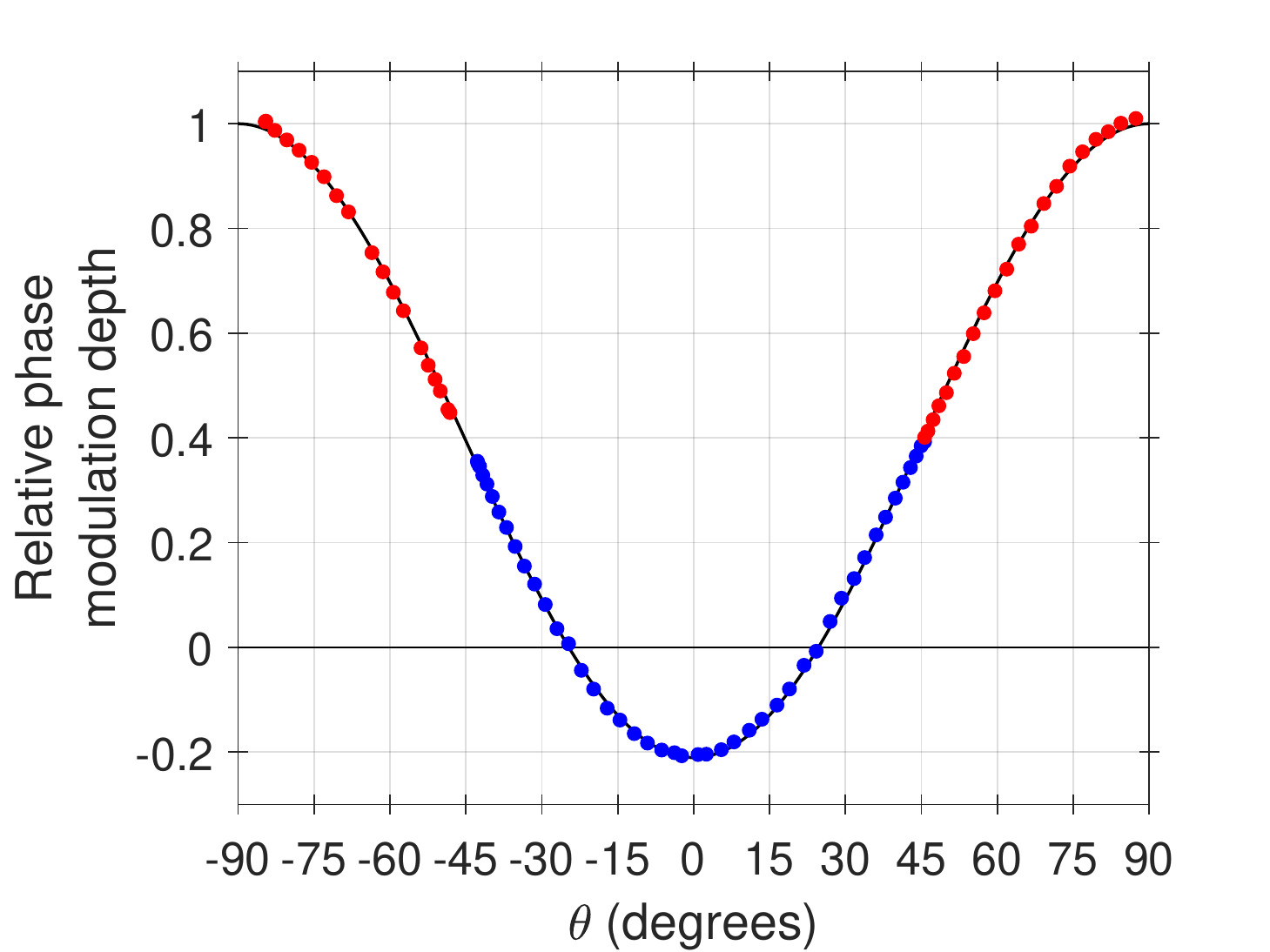} 
\caption{Measurements of the relative phase modulation depth (dots) are plotted along with the fitted theory curve (black line) for elliptical polarizations with $0^\circ \leq \theta \leq 45^\circ$ (blue dots) and $45^\circ \leq \theta \leq 90^\circ$ (red dots). $K = 300\,\mathrm{keV}$ in both cases. }
\label{sm:fig:elliptical}
\end{figure}

Equation \xeqn{eqn:phaserel}~ predicts that the spatial phase modulation profile should be independent of the EM standing wave's ellipticity angle, $\epsilon$, as defined in Eq. \xeqn{eqn:vectorpotential}~. To verify this prediction, the input half-wave plate was replaced with a quarter-wave plate. Rotating the quarter-wave plate then changed the polarization from either horizontal linear or vertical linear (depending on the linear polarization set before the quarter-wave plate) to left- or right-hand circular polarization. Intermediate angles of the quarter-wave plate generated elliptical polarizations. Therefore, $\epsilon$ is a function of the polarization angle $\theta$ in this configuration. Since the polarimeter used in this experiment was only capable of measuring $\theta$ (e.g. $\theta = 0^\circ$ for vertical linear polarization, $\theta = 45^\circ$ for circular polarization), the independence of the phase modulation depth on $\epsilon$ can only be inferred by comparing the measured values of the normalized phase modulation depth to those predicted by Eq. \xeqn{eqn:rho}~. This data is shown in Fig. \ref{sm:fig:elliptical} for two data sets---one taken with an elliptical polarization varying between horizontal linear and circular (red dots), and one taken with an elliptical polarization varying between vertical linear and circular (blue dots). Both data sets were taken at $K = 300\,\mathrm{keV}$. The measurements agree well with the $\epsilon$-independent theory.

\subsection{Observation of the phase profile at the relativistic reversal angle} 
When $\theta = \theta_r$, the electron beam acquires the spatial phase modulation profile $\phi\left(x,y\right) = - \phi_0\left(y\right) / 2$. Though such a phase profile does not cause electron diffraction along the laser axis, it can be used to implement phase contrast electron microscopy, where the relative phase shift between components of the electron wave function passing through and around the phase profile converts phase modulation in the electron wave function at the object plane to amplitude modulation in the image plane \xcite{ucb.b1269885920070101}~.

The Ronchigram imaging method used in this work is insensitive to the low spatial frequencies (relative to $1/\lambda_L$) which exist in the $- \phi_0\left(y\right) / 2$ phase shift profile. To qualitatively establish that the phase shift does not entirely disappear at the relativistic reversal angle, the electron microscope optics were aligned such that the diffraction plane and laser beam axis were coincident. In this configuration, the spatial phase profile imparted by the laser manifests in the contrast transfer function of the microscope and so appears in the Fourier transform of the image of a weak-phase sample \xcite{schwartz_laser_2018}~. 

\begin{figure}[H]
\centering
\includegraphics[width = 5in]{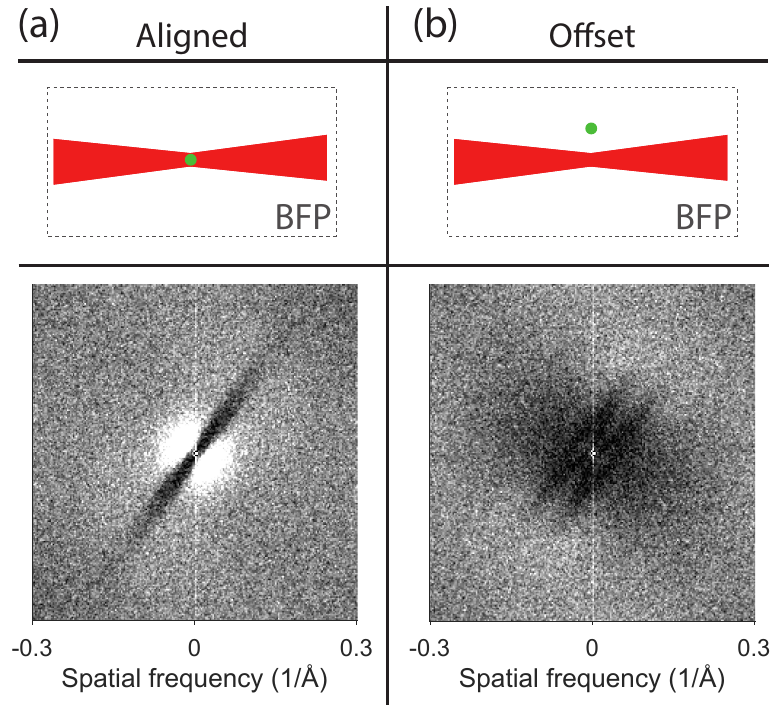} 
\caption{Laser phase shift profile when $\theta = \theta_r$, as observed through its effect on the Fourier transform of  the image of a weak phase sample. \textbf{a)} Center of electron diffraction plane (or back focal plane, BFP) aligned to the center of the laser beam. \textbf{b)} center of electron diffraction plane aligned off the side of the laser beam. This data shows that the phase shift does not disappear entirely when $\theta = \theta_r$, and the smooth Gaussian profile of the phase shift remains.}
\label{sm:fig:thincarbon}
\end{figure}

Figure \ref{sm:fig:thincarbon} shows Fourier transforms of the images of a thin carbon film sample under two electron beam alignment conditions. In both cases, $\theta = \theta_r$. In panel (a), the center of the electron beam diffraction pattern (green dot) was aligned to intersect the center of the laser beam. The profile of the laser-induced phase shift does indeed appear in the Fourier transform, with the laser axis running from the lower-left to upper-right. In panel (b), the center of the electron beam diffraction pattern was aligned off to the side of the laser beam. In this case, the laser-induced phase profile shows up in both the positive and negative frequency components of the image, as expected when the phase shift profile is asymmetric around the center of the diffraction pattern \xcite{ucb.b1269885920070101}~.

\subsection{Phase contrast imaging}
\begin{figure}[H]
\centering
\includegraphics[width = 5in]{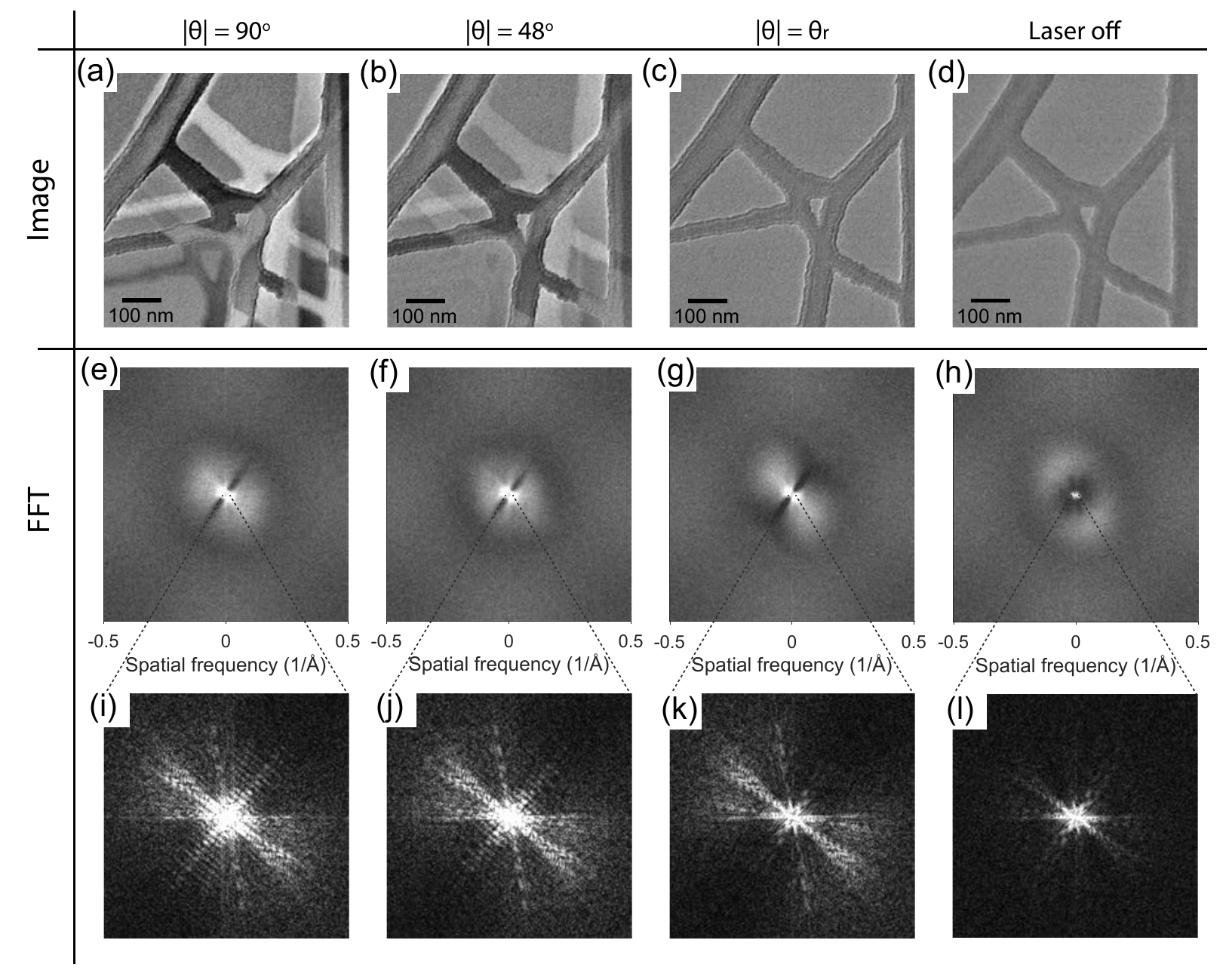} 
\caption{Laser phase contrast electron microscopy: ghost image artifacts and phase contrast as a function of polarization angle. \textbf{a-c)} Images of a lacey carbon sample using different polarization angles (annotated, top). \textbf{d)} Comparison with the laser off. Ghost artifacts are present when the laser is on and the polarization angle is not equal to the relativistic reversal angle $\theta_r$. \textbf{e-h)} Fourier transforms of panels (a-d), respectively. Phase contrast enhancement for all polarization angles is seen as increased power spectral density at low spatial frequencies, compared to the ``laser off" case (panel (h)). \textbf{i-l)} Detail of the lowest spatial frequencies of the Fourier transforms in panels (e-h), respectively. In panels (e) and (f), the standing wave of the laser beam is evident. At the relativistic reversal angle (panel (k)), the standing wave structure disappears from the Fourier transform. }
\label{sm:fig:ghosts}
\end{figure}

When the EM standing wave is used as a phase plate for phase contrast electron microscopy \xcite{schwartz_laser_2018}~, the relativistic polarization dependence of the potential can be used to control the spatial profile of the phase shift applied by the phase plate. When a standing wave structure exists in the phase shift, electrons diffract, causing ``ghost" image artifacts to appear in micrographs of strong phase objects \xcite{Schwartz:17}~. An example of such ghost artifacts is shown in the comparison between Fig. \ref{sm:fig:ghosts}(d) and Fig. \ref{sm:fig:ghosts}(a): the ``Laser off" image shows a lacey carbon sample with no laser phase plate present, while the ``$\left|\theta\right|=90^\circ$" panel shows the same region of the sample but imaged with the laser phase plate operated with a standing wave structure present in the phase shift. Ghost artifacts appear to the lower-left and upper-right of the primary object image (aligned with the standing wave axis). However, as the polarization is rotated to $\left|\theta\right|<90^\circ$, the phase shift modulation depth decreases, resulting in less electron diffraction and weaker ghost artifacts (Fig. \ref{sm:fig:ghosts}(b)). At the relativistic reversal angle $\theta_r$, the electron diffraction is eliminated entirely, and the ghost artifacts disappear (Fig. \ref{sm:fig:ghosts}(c)). However, the phase shift profile $\phi\left(x,y\right) = - \phi_0\left(y\right)/2$ still remains to provide phase contrast enhancement in the image (compare Fig. \ref{sm:fig:ghosts}(c) and Fig. \ref{sm:fig:ghosts}(d)---the former image has higher contrast). The remaining contrast enhancement is evident as increased low spatial frequency power spectral density in the Fourier transforms of the images (Fig. \ref{sm:fig:ghosts}(e-h), detail in Fig. \ref{sm:fig:ghosts}(i-l)).
